\documentstyle[12pt]{article}
\def\c{\cite}
\def\dsp{\displaystyle}
\def\txt{\textstyle}

\def\ee{\end{equation}}
\def\be{\begin{equation}}
\def\ep{\hspace*{\fill}$\Box$}
\def\A{{\cal A}}

\def\B{{\cal B}}
\def\ca{{}^*{\rm CAR}}
\def\con{{\rm const.}}
\def\df{:=}
\def\D{\partial}
\def\dip{\otimes}
\def\dis{\oplus}
\def\dl{d^2 l}
\def\e{\epsilon}

\def\F{{\cal F}}
\def\Fa{\widehat{{\rm CCR}}}
\def\H{{\cal H}}
\def\HI{{\cal H}_{IR}}
\def\hH{\hat{{\cal H}}}
\def\Hf{\hat{{\cal H}}_{[f]}}
\def\I{{\bf 1}}
\def\K{{\cal K}}
\def\Lf{L^f}
\def\La{\hat{L}^f}
\def\T{{\cal T}}
\def\U{{\cal U}}
\def\hU{\hat{U}}
\def\hW{\hat{W}}
\def\hpi{\hat{\pi}}
\def\hO{\hat{\Omega}}

\def\mo{\bmod 2\pi/e}
\def\n{\nabla}
\def\p{\psi}
\def\ph{\varphi}
\def\s{\!\cdot\!}
\def\ti{\tilde}
\def\tu{\tilde{u}}
\def\tV{\tilde{\dot{V}}}
\def\V{\dot{V}}
\def\w{\omega}
\def\W{\Omega}
\def\Ci{{\cal C}^\infty}

\def\CIR{{\cal C}^\infty_{IR}}

\def\R{{\rm \bf R}}
\def\Z{{\rm \bf Z}}

\def\C{{\rm \bf C}}

\def\lb{\left[}
\def\rb{\right]}
\def\lp{\left(}
\def\rp{\right)}
\def\a{\alpha}

\def\b{\beta}\def\dl{d^2 l}
\def\k{\kappa}

\def\g{\gamma}

\def\d{\delta}
\def\m{d\mu}

\def\la{\lambda}
\def\ov{\overline}

\newtheorem{defin}{Definition}
\newtheorem{pr}{Proposition}
\newtheorem{lem}[pr]{Lemma}
\newtheorem{col}[pr]{Corollary}
\newtheorem{theor}[pr]{Theorem}

\title{Semidirect product of CCR and CAR algebras and asymptotic
states in quantum electrodynamics}
\author{${\rm A{\scriptstyle NDRZEJ}~H{\scriptstyle ERDEGEN}}$
\thanks{e-mail: herdegen@thrisc.if.uj.edu.pl}\\
{\it Institute of Physics, Jagiellonian University,}\\ 
{\it Reymonta 4, 30-059 Cracow, Poland}}
\date{}
\begin{document}
\maketitle

\begin{abstract}
\noindent

A $C^*$-algebra containing the CCR and CAR algebras as its
subalgebras and naturally described as the semidirect product of
these algebras is discussed. A particular example of this
structure is considered as a model for the algebra of asymptotic
fields in quantum electrodynamics, in which 
Gauss' law is respected. The appearence in this algebra of a
phase variable related to electromagnetic potential leads to the 
universal charge quantization. Translationally covariant
representations of this algebra with energy-momentum spectrum in the
future lightcone are investigated. It is shown that vacuum
representations are necessarily nonregular with respect to 
total electromagnetic field. However, a class of
translationally covariant, irreducible representations is
constructed excplicitly, which remain as close as possible to the
vacuum, but are regular at the same time. The spectrum of
energy-momentum fills the whole future lightcone, but there are no
vectors with energy-momentum lying on a mass hyperboloid or in the
origin. 
\end{abstract}
\vfill
\eject

\begin{sloppypar}
\renewcommand{\thesection}{\Roman{section}}
\renewcommand{\theequation}{\arabic{section}.\arabic{equation}}
\renewcommand{\thepr}{\arabic{section}.\arabic{pr}}
\renewcommand{\thedefin}{\arabic{section}.\arabic{defin}}
\setcounter{equation}{0}
\setcounter{pr}{0}
\setcounter{defin}{0}

\section{\hspace*{-.7cm}. Introduction and summary}
\label{int}

In this paper we continue the study of the algebra of asymptotic
fields in quantum electrodynamics, in the framework developed earlier
in \c{her96} (and motivated by the classical asymptotic structure
discussed in \c{her95}). However, the present work is self-contained:
the main results of Ref.\c{her95} are recalled and the construction
of Ref.\c{her96} is restated here in a modified form. 

The existence of, and the algebraic relations between the asymptotic
(``in'' or ``out'') observables and fields in quantum electrodynamics
is a question of great physical interest. In the asymptotic limit, on
the one hand, the details and full complication of the dynamics
should lose their importance. On the other hand, the consequences of
Gauss' law and of the long-range character of the electromagnetic
interaction must survive. Suppose a closed algebra of asymptotic
fields in quantum electrodynamics may be constructed. This should
imply, at least, that (some) states over this algebra are approached
in the asymptotic limit by the expectation values of actual fields in
physical states.  Therefore, the infrared and charged structure of
the full theory should be encoded in the asymptotic algebra, and
physical insight into this structure may be gained by investigating
representations of this algebra.

Investigations into the infrared structure and asymptotic fields of
electrodynamics have a long history, see Refs.\c{jr}, \c{ms} and
\c{haa} for a review. They have led to the discovery of such
structures and effects as superselection sectors of the local
observables, the infraparticle problem, and spontaneous breaking of
the Lorentz symmetry. However, a clear formulation of a closed
algebraic structure of asymptotic fields has not been achieved,
although there do exist various partial answers to this problem, with
varying balance of mathematical rigour on the one hand and physical
concreteness on the other (the asymptotic dynamics of Kulish and
Faddeev \c{kf} and Zwanziger \c{zw}, the dressed electron states of
Fr\"ohlich \c{fr}, the asymptotic electromagnetic fields \c{bu} and
particle weights \c{bw} of Buchholz). The difficulties are twofold.
First, we do not have a complete, mathematically sound formulation
of QED. Second, the complete asymptotic separation of matter and
electromagnetic field may not be expected. These difficulties are not
of purely technical nature. The physical factor playing the decisive
role in the infrared structure of electrodynamics is the presence of
constraints, the Gauss' law. It is not clear at all, in our opinion,
to what degree the experience gained by the quantization of simpler,
unconstrained and short-range interactions may be taken over to the
formulation of quantum electrodynamics. In fact, the analysis of
Refs.\c{kf} and \c{fr} shows that the usual canonical quantization on
hypersurfaces of constant, finite time leads to the asymptotic
evolution which in the limit takes the states out of the space in
which the theory is defined for finite times.  This may be understood
as an indication that this method of obtaining a quantum theory may
not be best suited for electrodynamics. Also, the localization
properties of observables which form the base of the axiomatic
algebraic approach to quantum field theory problems still need
justification in the case of electrodynamics, and may seem somewhat
artificial in the version considered there (localization in spacelike
cones).

The above remarks should not be understood as an attempt to question
{\em a priori} the standard wisdom on the subject. We would like,
rather, to indicate that the problem of the algebraic structure of
quantum electrodynamics may still be validly regarded as open. This
view is also supported by the failure of the standard electrodynamics
to explain such striking and universal physical fact as the observed
spectrum of charge \c{sta}.

In the present work we follow the approach of Refs.\c{her96} and
\c{her95} to formulate and investigate a concrete $C^*$-algebra of
asymptotic fields in electrodynamics. The basic idea of the approach
consists of interchanging the order of ``quantization'' and
``asymptotic limit''. (Quantization of free electromagnetic field
along these lines has been discussed in Refs.\c{bras} and \c{asht}.
Our approach develops the ideas of Ref.\c{asht} and extends the
program to include charged fields.) In
Ref.\c{her95} the asymptotic structure of the classical field
electrodynamics (the Maxwell-Dirac system) has been discussed. In
Ref.\c{her96} then a class of models of the asymptotic algebra has
been obtained by the quantization of this classical structure by the
correspondence principle. The Gauss' law is naturally implemented on
the algebraical level. Now we develop the approach further in the
following ways. \\ 
(i) We restrict attention to the technically
simplest of the models introduced in Ref.\c{her96}, the one in which
only the Coulomb field is undetachable from a particle (the
possibility (b) of Sec.VI in Ref.\c{her96}). While not rejecting
other possibilities at this stage, we observe that this model is also
the one following most naturally by the quantization of the classical
structure. In this case the construction of Ref.\c{her96} may be
simplified to yield the asymptotic algebra in the form of a
particular instant of a semidirect product of the CCR (canonical
commutation relations) and CAR (canonical anticommutation relations)
algebras. The definition and the discussion of some mathematical
aspects of this structure comprise the self-contained Section
\ref{semi} of the paper.\\ 
(ii) In the course of quantization we
bring into play a new factor, which is present at the classical level
in Ref.\c{her95}, but has not been properly taken into account in the
quantization procedure in Ref.\c{her96}. It has been shown in
Ref.\c{her95} that one of the asymptotic variables has a natural
interpretation as a phase variable. The quantization of the classical
structure which properly respects the phase character of this
variable leads unambiguously to the quantization of the physical
charge spectrum in units of elementary charge. As the phase variable
is connected with the free electromagnetic potential, this
quantization law is universal: it has to be respected by any carriers
of charge. (For similar reasoning, but with a different
identification of the phase variable, see the works by Staruszkiewicz
\c{sta}.) The classical asymptotic structure, its quantization and
the resulting algebra are discussed in Sec.\ref{asym}, including the
action of the Poincar\'e group on the algebra and the identification
of observables. \\
(iii) Section \ref{grep} contains some general results on physically
admissible representations of the asymptotic algebra. It is argued
that the representations should be regular with respect to all Weyl
operators, as otherwise the Coulomb field part of the total field is
lost. In representations satisfying Borchers' criterion (the
spectrum of energy-momentum in the future lightcone) the Gauss
constraint, which is hidden in the commutation relations on the
abstract algebraic level, is shown to be recovered in the functional
form. \\
(iv) More special representations are investigated in Sec.\ref{vreg}
with the use of technical tools discussed in Sec.\ref{auw} (some
technical material is shifted to the Appendix). All vacuum representations
are shown to be nonregular with respect to the Weyl operators with
infrared-singular test functions, which is a handicap as explained
earlier. Also, superselection sectors with respect to regular
operators are distinguished by the {\em free} field spacelike
asymptotic, which is not what one would expect in a physical state
\c{buch}. There exists a Poincar\'e-invariant vacuum on the field
algebra (sectors transform into each other under Lorentz
transformations), whose GNS representation space contains charge-one
vector states with the energy-momentum on the mass hyperboloid (no
infraparticle problem in this vacuum).  A class of representations
satisfying Borchers' criterion is constructed, which remain as close
as possible to this representation, but are regular at the same time.
The nonexistence of charged states on the mass hyperboloid follows
here from nonexistence of a vacuum vector state. It is not known at
present whether there do exist regular Poincar\'e covariant
representations satisfying Borchers' criterion (standard arguments 
\c{fms,buch} do not apply here).

\setcounter{equation}{0}
\setcounter{pr}{0}
\setcounter{defin}{0}

\section{\hspace*{-.7cm}. Semidirect product of CCR and CAR algebras}
\label{semi}

In this section we discuss a generalization of the direct product of
CCR and CAR algebras\footnote{The resulting algebra and some
of the statements in the present section may be obtained, as
pointed out to me by H.\,Grundling, by the application of the general
theory of twisted crossed products of $C^*$-algebras by groups
\c{para}.}. We also identify a class of representations of the
resulting $C^*$-algebra. 

Suppose we are given the following constructs:\\
(i) a complex Hilbert space $\K$ and the *-algebra $\ca$ generated
by the elements $a(f)$ depending antilinearly on $f\in\K$ and by the
identity $E$, according to the CAR relations (eqs. (\ref{f}) below);\\
(ii) an Abelian group $X$ (with the additive notation of the group
multiplication) equipped with a symplectic form $X\times X\ni (x,y)
\to \{ x,y\} \in \R$;\\
(iii) a representation of the group $X$ in the automorphism group of
the *-algebra $\ca$
$$
X\ni x\to\b_x\in\mbox{Aut}\ca\, ,~~\b_x\b_y =\b_{x+y}\, .
$$

We consider the *-algebra $\B$ generated by the elements $a(f)~
(f\in\K)$, $W(x)~(x\in X)$ and the identity $E$ according to the
relations: 
\be
W(x)^* = W(-x)\, ,~W(0)=E\, ,~W(x)W(y)=e^{\txt -\frac{i}{2}\{ x,y\}}
W(x+y)\, ,
\label{w}
\ee
\be
[a(f), a(g)]_+ =0\, ,~~~[a(f), a(g)^*]_+ = (f, g) E\, ,
\label{f}
\ee
\be
W(x) C = \b_x(C)\, W(x)~~\forall C\in\ca\, .
\label{sd}
\ee
With the use of (\ref{sd}) every element of $\B$ may be given the
form $\dsp\sum_{i=1}^n C_i W(x_i)$, where $C_i\in\ca$. 
\begin{pr} \hspace*{\fill}\\
(i) If $x_i\neq x_k$ for $i\neq k$ and $C_i\neq 0$ in $\ca$ then 
$\dsp\sum_{i=1}^n C_i W(x_i) \neq 0$. Every nonzero element of $\B$
is uniquely represented in this way. \\
(ii) The *-algebra $\B$ has a faithful representation by bounded
operators in a Hilbert space. 
\label{repb}
\end{pr}
Proof. Let $\pi_f$ be a representation of $\ca$ in a Hilbert space
$\H$ (e.g. the Fock representation). Let $\pi(W(y))$ and $\pi(C)$ be
operators acting in the direct sum Hilbert space 
$\dsp\bigoplus_{x \in X} \H_x$, $\H_x=\H$, defined by 
$$
\lb\pi(W(y))\p\rb_x = e^{\txt \frac{i}{2}\{ x,y\} }
\p_{x-y}\, ,~~\lb \pi(C)\p\rb_x = \pi_f(\b_{-x}C)\p_x\, ,
$$
where $\dsp\p =\bigoplus_x \p_x$. One
easily shows that these bounded operators satisfy the relations
(\ref{w}-\ref{sd}) and part (i) of the proposition. This implies
part (ii).
\ep

This representation defines a $C^*$-norm on the algebra $\B$. All
$C^*$-norms on $\B$ are jointly bounded, as for each such norm $p$
there is $p\lp W(x)\rp=1$ and $p\lp a(f)\rp=\| f\|_\K$. Therefore,
the set of $C^*$-norms contains the maximal element defined by
$\dsp\| A\|\df\sup_p p(A)$ \c{dixc}.
\begin{defin}
The field algebra $(\F,\| .\|)$ is the $C^*$-completion of the 
*-algebra $\B$ in the norm $\| .\|$.
\label{field}
\end{defin}
Remarks.\\
(i) The elements $a(f)$ (resp. $W(x)$) generate a $C^*$-sub\-al\-gebra of
$\F$ which will be called the CAR (resp. the CCR, or the Weyl)
algebra in this article.\\
(ii) The construction would be slightly more general if the CAR
algebra with its auto\-morphisms instead of $\ca$ and $\beta_x$ were
used. This, however, would not be convenient for our purposes.
\begin{col}{}\hspace*{\fill}\\
(i) Every Hilbert space representation of $\B$ is given in terms of
bounded operators and extends to a representation of $\F$.\\
(ii) Every positive linear functional on $\B$ defines via the GNS
construction a Hilbert space representation of $\B$ and extends to a
positive linear functional on $\F$.\\
(iii) Every *-automorphism of $\B$ extends to a *-automorphism of
$\F$.
\label{repF}
\end{col}
We omit a simple proof. The corollary establishes the 1:1
corespondence between the positive functionals and representations of
$\F$ on the one hand, and of $\B$ on the other hand. It also gives a 
class of automorphisms of $\F$ in simple terms. We shall repeatedly 
(and tacitly, mostly) take advantage of these simplifications in what
follows.

An important class of automorphisms of $\B$ is given as follows. Let
$T$ be a symplectic additive mapping in $X$:
\be
T(x+y)=Tx+Ty\, ,~~ \{ Tx, Ty\} =\{ x,y\}
\label{bo1}
\ee
Let, further, $\tau$ be an automorphism of $\ca$ satisfying for each
$x \in X$
\be
\tau \beta_x=\beta_{Tx}\tau
\label{bo2}
\ee
Define $\tau\lp W(x)\rp=W(Tx)$. Then $\tau$ extends to an
automorphism of $\B$ (and $\F$). This is a Bogoliubov transformation
when restricted to the Weyl algebra.

In the following sections the constructions outlined above will be
needed in a special case. Both $\beta_x$ and $\tau$ will be
Bogoliubov transformations given by
\be
\beta_x a(f)=a(S_xf)\, ,~~ \tau a(f)=a(Rf)\, ,
\label{aut}
\ee
where $S_x$ is a unitary representation of $X$ in $\K$ and $R$ is a
unitary operator in $\K$. The condition (\ref{bo2}) now takes the
form 
\be
RS_x=S_{Tx}R\, .
\label{bo3}
\ee

In the remaining part of this section we introduce a particular class
of representations of $\B$. We pose, namely, the following question.
Suppose $\w_f$ and $\w_r$ are states on the CAR and the Weyl algebra,
respectively. When does  $\w\lp CW(x)\rp= \w_f(C)\,\w_r\lp W(x)\rp$
define a state on $\F$? The hermiticity of $\w$ implies that
$\lp\w_f(\beta_x C)-\w_f(C)\rp\,\w_r\lp W(x)\rp =0$ for every $C \in
\mbox{CAR}$ and $x\in X$.
This condition is satisfied, in particular, in each of the following
two cases:
\be
\w_f(\b_x C)=\w_f(C)~~\mbox{for all}~~x \in X\, ,
~~ C\in\mbox{CAR}\, , 
\label{cf}
\ee
\be
 \w_r\lp W(x)\rp =0~~\mbox{if}~~\beta_x\neq \mbox{id}\, .
\label{cr}
\ee
One easily shows that in each of these cases $\w$ is also positive on
$\B$, hence it is a state on $\F$. 
We generalize this construction and write down the explicit prescription
for the resulting representation in terms of $\pi_r$ and $\pi_f$ in
the following statement. 
\begin{pr}
Suppose $\pi_f$ and $\pi_r$ are representations of the {\rm CAR} 
algebra and the Weyl algebra in Hilbert spaces $\H_f$ and $\H_r$, 
respectively. Define operators $\pi(A)$, acting in the space 
$\H_f \dip \H_r$, in the following two cases:\\
(i) $\pi_f$ has a cyclic vector $\W_f$, such that the expectation
values in that state satisfy (\ref{cf}); then
\be
\begin{array}{@{}l}
\dsp \pi (C) \df\pi_f(C) \dip\I_r ,\\
\dsp \pi\lp W(x) \rp\, \lb \pi_f(B)\W_f \dip\ph\rb
\df \pi_f(\b_x B) \W_f \dip \pi_r \lp W(x) \rp \ph \, .
\end{array}
\label{repf}
\ee
(ii) $\pi_r$ has a cyclic vector $\W_r$, such that the expectation
values in that state satisfy (\ref{cr}); then
\be
\begin{array}{@{}l}
\dsp \pi \lp W(x) \rp \df\I_f \dip\pi_r \lp W(x)\rp\, ,\\
\dsp \pi(C)\, \lb \p\dip\pi_r\lp W(y)\rp \W_r \rb\df
\pi_f(\b_{-y} C) \p \dip \pi_r\lp W(y)\rp \W_r\, ;
\end{array}
\label{repr}
\ee
Then $\pi$ defines a representation of the algebra $\B$ (and $\F$) in
each of the two cases.
\label{srep}
\end{pr}
Proof. Part (i) is immediately verified if one observes that 
$\pi(W(x))=U_x\dip \pi_r(W(x))$, where $U_x$ canonically implement 
$\beta_x$ in the representation generated by $\w_f$. To prove (ii)
one has to show that $\pi (C)$ in (\ref{repr}) is well defined as a 
linear operator. Once this is established, the verification of
the algebraic properties is a simple exercise, which we omit. 
Let $\sum_{i=1}^n \p_i\dip\pi_r \lp W(x_i) \rp \W_r  = 0$. 
Without loss of generality
$\b_{x_i} = \b_{x_k} (i,k = 1,\dots,n)$, as otherwise the vectors are
orthogonal. Then $\sum_i  \pi_f(\b_{-x_i}C) \p_i\dip\pi_r \lp
W(x_i) \rp \W_r = 
\pi_f(\b_{-x_1}C)\dip\I_r \sum_{i=1}^n  \p_i \dip\pi_r \lp W(x_i) \rp
\W_r = 0$,  which justifies the definition of $\pi(C)$.   \ep

The following result will be needed later. 
\begin{pr}
Let $\tau$ be an automorphism of $\F$ satisfying 
(\ref{bo1},\ref{bo2}), and
let $\pi$ be a representation of $\F$ of the type (\ref{repf}). The
following two conditions are equivalent.\\
(i) The symmetry $\tau$ is implementable in the representation $\pi$
by a unitary operator leaving the subspace $\W_f\dip\H_r$ invariant:
\be
\pi(\tau A)=U\pi(A)U^*~~~\forall A\in\F\, ,
\label{tu1}
\ee
\be
U\lp\W_f\dip\H_r\rp \subset \W_f\dip\H_r\, .
\label{tu2}
\ee
(ii) The state $\w_f$ is invariant under $\tau$
\be
\w_f(\tau C)=\w_f(C)~~~\forall C\in {\rm CAR}\, ,
\label{tu3}
\ee
and $\tau$ is implementable in the representation $\pi_r$
\be
\pi_r(\tau A)=U_r\pi_r(A)U_r^*~~~\forall A\in {\rm CCR}\, .
\label{tu4}
\ee
If these conditions are satisfied, there is a $1:1$ correspondence
between operators $U$ and $U_r$ given by 
\be
U=U_f\dip U_r\, ,~~\mbox{where}~~U_f\pi_f(C)\W_f
\df\pi_f(\tau C)\W_f\, .
\label{tu5}
\ee
If, moreover, $G\ni g\rightarrow\tau_g\in {\rm Aut}\F$ is a symmetry
group then $U(g)$ is a representation of $G$ iff $U_r(g)$ is. 
\label{tu}
\end{pr}
Proof. (i) $\Rightarrow$ (ii). If $U$ satisfies (\ref{tu2}), then the
equation $U\lp\W_f\dip\ph\rp =\\ 
\W_f\dip U_r\ph$ defines a unitary operator $U_r$. 
Then, by (\ref{repf}) and (\ref{tu1}), \\
$\W_f\dip U_r\pi_r(W(x))\ph = U\pi(W(x))(\W_f\dip \ph)= 
\W_f\dip \pi_r(\tau W(x)) U_r\ph$, which is equivalent to
(\ref{tu4}). Moreover, again by (\ref{repf}) and (\ref{tu1}), 
$U(\pi_f(C)\W_f\dip \ph)=U\pi(C)(\W_f\dip\ph)= \pi_f(\tau C)\W_f\dip
U_r\ph$, hence $U$ has the form (\ref{tu5}). Finally, 
$\w_f(\tau C)= (\W_f, \pi_f(\tau C)\W_f)= \w_f(C)$. \\
(ii) $\Rightarrow$ (i). Choose $U$ as in (\ref{tu5}). Then
(\ref{tu1}) for $A\in\mbox{CAR}$ is obvious from (\ref{repf}).
Moreover, by (\ref{repf}) and (\ref{bo2}), 
\begin{eqnarray*}
&&U\pi(W(x))U^*\lp \pi_f(C)\W_f\dip\ph\rp 
= \pi_f(\tau\b_x\tau^{-1} C)\W_f\dip U_r\pi_r(W(x))U_r^*\ph\\
&&= \pi_f(\b_{Tx}C)\W_f\dip\pi_r(W(Tx))\ph 
= \pi(W(Tx))\lp \pi_f(C)\W_f\dip\ph\rp\, .
\end{eqnarray*}
If $\tau_g$ is a symmetry group then $U_f(g)$ is a representation of
$G$, which implies the last equivalence of the proposition.  \ep \\
A similar proposition holds for the representations of the type 
(\ref{repr}). 

\setcounter{equation}{0}
\setcounter{pr}{0}
\setcounter{defin}{0}

\section{\hspace*{-.7cm}. The asymptotic field algebra}
\label{asym}

We turn in this section to the proper task of this article, the
investigation of an asymptotic algebra of fields for charged
particles in interaction with an electromagnetic field, as outlined
in the Introduction. For this purpose it is necessary to review the
asymptotic structure of the classical field electrodynamics (the
Maxwell-Dirac system) discussed in Ref.\c{her95} (see also
Ref.\c{asht}).  This paper contains, more precisely, rigorous results
for both external field problems and supplies plausibility arguments
for the persistence of the resulting structures in the fully
interacting theory. (For recent rigorous results on the dynamics of
the classical Maxwell-Dirac system obtained by the adaptation of a
modified Dollard method \c{dol} see Ref.\c{fla}; our approach is
different.) Here we only briefly sketch the results without bothering
about regularity assumptions or the exact sense of limits, with the
purpose of merely identifying the asymptotic variables. For more
details we refer the reader to \c{her95}. On the other hand it should
be clear that quantization by the correspondence principle is a
heuristic procedure itself, so there is no need for excessive
formalization of this point.

Due to the difference in propagation of the electromagnetic field on
the one hand, and massive fields on the other, the natural asymptotic
directions are also different in the two cases, lightlike in the
first and timelike in the second case, respectively. Consider the
electromagnetic field first. Let $l_a$ be a null, future-pointing
vector ($a$, $b$, etc. are spacetime indices). Then the leading
asymptotic term in this direction of the electromagnetic potential in
a Lorentz gauge $A_a(x)$ is given by
\be
A_a(x+Rl)\sim \frac{1}{R} V_a(x\s l,l)~~~\mbox{for}~~R\to\infty\, .
\label{as1}
\ee
where $x$ is any spacetime point, $V_a(s,l)$ is a real,
spacetime-vector-valued function of a real variable $s$ and a null
vector $l$, and $x\s l$ denotes the scalar product with signature
$(+,-,-,-)$. Different gauges yield $V$'s differing by the
transformation $V_a(s,l)\to V_a(s,l)+l_a \a(s,l)$, which will also be
referred to as gauge transformation, but determine the same
electromagnetic field asymptotic 
\be
F_{ab}(x+Rl)\sim \frac{1}{R}\lp l_a\V_b(x\s l,l)-l_b\V_a(x\s l,l)\rp 
~~~\mbox{for}~~R\to\infty\, ,
\label{as2}
\ee
where the dot over $V$ denotes differentiation with respect to $s$. 
The functions $V_a(s,l)$ are homogeneous of degree $-1$
\be
V_a(\mu s,\mu l) = \mu^{-1} V_a(s, l)\, ,~~\mu>0\, ,
\label{hom}
\ee
satisfy
\be
l\s V(s, l) = Q\, ,
\label{ch}
\ee
where $Q$ is the total charge of the system as measured by the
inegrated electric flux in spacial infinity, and tend to the limit
functions $V_a(\pm\infty,l)$ for 
$s\to\pm\infty$. The homogeneity implies that $V_a(s,l)$ is
determined by its values for $l$'s on a manifold cutting each null
direction once. In each Minkowski frame, if $l$'s are scaled to
satisfy $t\s l=1$ (where $t$ is the timelike basis vector of the
frame), then $\V_a(s,l)$ falls off componentwise for $|s|\to\infty$
at least as $|s|^{-(1+\e)}$, for some $\e>0$. Differentiations with
respect to independent variables in the null vector $l$ (i.e., 
differentiations in directions tangent to the lightcone) are
conveniently carried out with the use of the operator $L_{ab}\df
l_a\D_b - l_b\D_a$, $\D_a\df\D/\D l^a$. With this notation the limit
functions $V_a(\pm\infty,l)$ satisfy the differential condition 
\be
L_{[ab}V_{c]}(\pm\infty,l)=0\, .
\label{eL}
\ee

The limit function $V_a(-\infty,l)$ has a clear physical meaning.
For any spacelike vector $y$ there is 
\be
\lim_{R\to\infty}R^2\, F_{ab} (x + Ry) 
= {1\over 2\pi}\int\lp l_a V_b (-\infty, l) - l_b V_a (-\infty, l)\rp
\delta'(y\s l)\, \dl\, , 
\label{spas}
\ee
where $\delta'(.)$ is the derivative of the Dirac delta-function and
$\dl$ is the Lorentz-invariant measure on the set of null directions
applicable to functions homogeneous in $l$ of degree $-2$. 
(If $l$'s are scaled to $t\s l=1$, then $\dl$ is the spherical angle
measure on the unit sphere in the hyperplane orthogonal to $t$.)
Therefore, $l\wedge V(-\infty,l)$ is responsible for the long-range
part of the electromagnetic field. The physical content of the
property (\ref{eL}) of $V(-\infty,l)$ is that the long-range field is
of purely electrical type. The physical meaning of the limit function
$V(+\infty,l)$ and its property (\ref{eL}) will become clear in the
sequel. 

Consider now the timelike asymptotic of the Dirac field, in the sense
of asymptotic behaviour for $\lambda\to\infty$ of $\psi(\la v)$,
where $v$ lies on the future part of the unit hyperboloid. In the
Dirac equation choose the electromagnetic potential in a local gauge
$A^{{\rm tr}}$, related (locally) to a Lorentz potential $A$ by
$A^{{\rm tr}}_a(x)= A_a(x)-\n_a S(x)$, with the condition that for
$x^2\to+\infty$, $x^0>0$ the leading term of $S(x)$ is $\ln\sqrt{x^2}
\, x\s A$. Then the leading asymptotic term of the Dirac field in
this gauge, $\p^{{\rm tr}}(x)=e^{-ieS(x)}\p(x)$, is 
\be
\psi^{{\rm tr}}(\la v)\sim -i\la^{-3/2} 
e^{\txt -i(m\la+\pi/4)\g\s v} f(v)~~{\rm for}~~\la\to\infty\, ,
\label{tas}
\ee
where $\g^a$ are the Dirac matrices. The function $f(v)$ is a
$\C^4$-valued function with finite norm squared 
$\dsp\int\ov{f(v)}\g\s vf(v)\, d\mu(v)$, where bar denotes the usual
Dirac conjugation and $d\mu(v)$ is the invariant measure $d^3v/v^0$.
The merit of the above asymptotics lies in its
simplicity and form-independence of the potential, as long as the
latter lies in the distinguished class. This does not contradict the
modified Dollard asymptotic dynamics \c{dol,kf,fla}, as here the
asymptotic sequence of hyperboloids rather than spacelike hyperplanes
is considered, and a special class of gauges is used. 

The asymptotic variables $V_a(s,l)$ and $f(v)$ have well-defined
transformation properties under the action of the Poincar\'e group.
The element $(x,A)$ of its universal covering group ($x$ is a
spacetime vector and $A\in SL(2,\C)$) transforms the Lorentz
(covariant) gauge fields by $A_a(y)\to\Lambda(A)_a{}^b
A_b(\Lambda^{-1}(y-x))$, $\p(y)\to S(A)\p(\Lambda^{-1}(y-x))$, where
$S(A)$ is the bispinor representation and $\Lambda(A)$ is the
representation of $SL(2,\C)$ in ${\cal L}_+^\uparrow$ (with the
notation $(\Lambda y)^a=\Lambda^a{}_by^b$). The field $\p^{{\rm
tr}}(y)$ transforms noncovariantly, but the noncovariant phase
phactors are shown to cancel out in the limit. The asymptotics
transform then by the representations of the Poincar\'e group 
\begin{eqnarray}
&&\lb T_{x,A} V\rb_a (s,l) 
= \Lambda(A)_a{}^b\, V_b(s-x\s l,\Lambda^{-1} l)\, ,
\label{poin1}\\
&&[R_{x,A}f](v) = e^{\txt imx\s v\g\s v}S(A)f(\Lambda^{-1}v)\, .
\label{poin2}
\end{eqnarray}
Let us introduce the following structures on the space of asymptotic
variables: the symplectic form
\be
\{ V_1, V_2\} = \frac{1}{4\pi}\int \lp \V_1\s V_2 - \V_2\s V_1 \rp
\! (s,l)\, ds\, \dl\, ,
\label{sympl}
\ee
and the scalar product
\be
(f_1,f_2) = \int\ov{f_1(v)}\g\s v f_2(v) \m(v)\, ,
\label{sp}
\ee
(both of them are well defined). Then $T_{x,A}$ is a symplectic
transformation, and $R_{x,A}$ a unitary one:
\be
\{ T_{x,A}V_1, T_{x,A}V_2\} =
\{ V_1, V_2\} ~,~~(R_{x,A}f_1,R_{x,A}f_2)=(f_1,f_2)\, .
\label{su}
\ee
The generators of these transformations defined by 
$$
T_{x,A}-\I\approx x^ar_a-\frac{1}{2}\w^{ab}n_{ab}\, ,~
R_{x,A}-\I\approx ix^ap_a-\frac{i}{2}\w^{ab}m_{ab}\, ,
$$
for infinitesimal $x^a$ and $\w^{ab}$, 
where $\Lambda^a{}_b\approx g^a_b+\w^a{}_b$, are 
\begin{eqnarray*}
&&(r_aV)_c(s,l)=-l_a\V_c(s,l)\, ,~(n_{ab}V)_c(s,l)
=-L_{ab}V_c(s,l)-g_{ca}V_b(s,l)+ g_{cb}V_a(s,l)\, ,\\
&&(p_af)(v)=mv_a\g\s vf(v)\, ,~(m_{ab}f)(v)
=\lp v_ai\d_b-v_bi\d_a +\frac{i}{4}[\g_a,\g_b]\rp f(v)\, , 
\end{eqnarray*}
where $\d_a$ is the derivative tangent to the hyperboloid, 
$\d_af(v)\df (\n_a-x_ax^c\n_c)f(x)\Big|_{x=v}$, and on the r.h.side
any extension of $f(v)$ to the local neighborhood of the hyperboloid
is used. 

The discussion of Ref.\c{her95} suggests that the asymptotic
variables $V(s,l)$ and $f(v)$ form a causally complete set,
in the sense that they determine the state of the system at any
spacetime point (this has not been proved in the present
approach, but cf. Ref.\c{fla}). The total energy-momentum and angular
momentum are shown to be sums of two terms, the first one describing
the respective quantity going out in timelike directions and
expressed in terms of $f(v)$ only, and the second one describing the
respective quantity going out in lightlike directions and expressed
in terms of $V_a(s,l)$ only. (For angular momentum this is, actually,
the natural and well-defined way for extending the definition of this
quantity to the infrared-singular case; the standard integral over a
Cauchy surface, as resulting from Noether theorem, is ill defined.)
Explicitly, they may be put into the form
\begin{eqnarray}
&&P_a=(f, p_af) +{\txt\frac{1}{2}}\{ V,r_aV\}\, ,
\label{enm}\\
&&M_{ab}=(f, m_{ab}f) +{\txt\frac{1}{2}}\{ V,n_{ab}V\}\, .
\label{anm}
\end{eqnarray}
(The first terms in these formulae are those of Eqs.(5.15) and (5.16)
in Ref.\c{her95}, while the second ones are the tensor forms of the
expressions (3.13) and (3.14) in the same reference.) 
Note that the local gauge freedom of the Dirac field is lost in the
asymptotic limit as defined in the present approach -- a change of
phase of $f(v)$ by a nonconstant function of $v$ spoils the form of
the angular momentum. The electromagnetic terms are separately gauge
invariant, although the symplectic form (\ref{sympl}) depends on
gauges of $V_{{\rm i}}{}_a(-\infty,l)-V_{{\rm i}}{}_a(+\infty,l)$,
${\rm i}=1,2$ (cf. below). 

The conditions of relativistic quantization for the quantum variables
$V^{{\rm q}}$ and $f^{{\rm q}}$ corresponding to the classical ones
are \c{sym} 
\be
\begin{array}{@{}ll}
\dsp [P_a,V^{{\rm q}}_c]=i(r_aV^{{\rm q}})_c\, ,
&\dsp [M_{ab},V^{{\rm q}}_c]=i(n_{ab}V^{{\rm q}})_c\, ,\\
\dsp [P_a,f^{{\rm q}}]=-p_af^{{\rm q}}\, , 
&\dsp [M_{ab},f^{{\rm q}}]=-m_{ab}f^{{\rm q}}\, ,
\end{array}
\ee
where in the generators also the respective quantum variables should
be substituted. Suppose, that the variables $V^{{\rm q}}$ and
$f^{{\rm q}}$ commute, $[V^{{\rm q}}, f^{{\rm q}}]=0$ and all
fundamental (anti)commutators are c-numbers. This
assumption remains in concord with the standard wisdom on canonical
quantization in local gauges (on which the derivation of both
variables is based) and fixes the quantization rules uniquely:
\be
\lb\{ V_1,V^{{\rm q}}\}, \{ V_2,V^{{\rm q}}\}\rb = 
i\{ V_1, V_2\}\, ,
\label{hq1}
\ee
\be
\lb (f_1,f^{{\rm q}}), (f_2,f^{{\rm q}})\rb_+=0\, ,~
\lb (f_1,f^{{\rm q}}), (f_2,f^{{\rm q}})^*\rb_+= (f_1, f_2)\, .
\label{hq2}
\ee

The above quantization relations must be considered as merely the
first step towards our aim, as up to now we have not taken into
account the constraints between the asymptotic variables. To remedy
this deficiency we return to the discussion of the classical
structure. The symplectic form (\ref{sympl}) is invariant under the
constant gauge transformation $V_a(s,l)\to V_a(s,l)+l_a\a(l)$. One
shows that with the appropriate choice of this gauge there is 
\be
V_a(+\infty,l)=\int n(v) V^e_a(v,l)\, \m(v)\, ,
\label{Gau}
\ee
where $n(v)=\ov{f(v)}\g\s v f(v)$ is the asymptotic density of
particles moving with velocity $v$ and $V_a^e(v,l)
=ev_a/v\s l$ is the null asymptotic (\ref{as1}) of the
Lorentz potential of the Coulomb field surrounding a particle with
charge $e$ moving with constant velocity $v$. Therefore, the above
relation is the implementation of the Gauss constraint on the space
of classical asymptotic variables. The relation (\ref{eL}) for this
limit function is now seen to be satisfied identically. Let,
furthermore, $A_a^{{\rm adv}}(x)$ be the advanced field of the
sources. It turns out that the asymptotic (\ref{as1}) of this field
is given by $V_a^{{\rm adv}}(s,l)=V_a(+\infty, l)$. Hence, the
asymptotic of the free outgoing field potential, standardly defined
by $A^{{\rm out}}=A-A^{{\rm adv}}$, is determined by 
$V_a^{{\rm out}}(s,l)=V_a(s,l)-V_a(+\infty,l)$. The ``out'' field is
recovered from its asymptotic by the formula
\be
A_b^{{\rm out}}(x)=-\frac{1}{2\pi}\int \V_b^{{\rm out}}(x\s l, l)
\, \dl\, .
\label{pot}
\ee
The connection of this formula with the Fourier representation 
\be
A_b^{{\rm out}}(x)=\frac{1}{\pi}\int a_b(k) \delta(k^2)\e(k^0) 
e^{-ix\cdot k}\, d^4k
\label{OF}
\ee
is supplied by the relation $a_b(\w l)=-\w^{-1}\tV_b^{{\rm
out}}(\w,l)$, where the following one-dimensional Fourier
transformation has been introduced
\be
\ti{h}(\w,l) = \frac{1}{2\pi} \int h(s,l)e^{\txt i\w s} ds\, .
\label{four2}
\ee
The limit function $V^{{\rm out}}(-\infty,l)
=-2\pi\tV^{{\rm out}}(0,l)$ describes the long-range
(infrared-singular) part of $A^{{\rm out}}$, the limit function at
$s\to\infty$ vanishes, and the charge in formula (\ref{ch}) is zero. 

The infrared characteristic $V^{{\rm out}}(-\infty,l)$ has a simple
representation, to become of importance below. Eqs (\ref{hom}),
(\ref{ch}) (with $Q=0$), and (\ref{eL}) satisfied by $V^{{\rm
out}}(-\infty,l)$ imply that there exists a homogeneous of degree $0$
function $\Phi(l)$ such that 
\be
l_aV^{{\rm out}}_b(-\infty,l)-l_bV^{{\rm out}}_a(-\infty,l) 
= L_{ab}\Phi(l)\, .
\label{elfi}
\ee
The function $\Phi(l)$ is determined up to an additive constant, but
one of the solutions is distinguished by being determined linearly
and Lorentz-covariantly by $V^{{\rm out}}_a(-\infty,l)$:
\be
\Phi_{V^{{\rm out}}}(l) = \frac{1}{4\pi}\int
\frac{l\s V^{{\rm out}}(-\infty,l')}{l\s l'}\,\dl'\, .
\label{FV}
\ee
(This explicit formula appears here for the first time; it may be
obtained by a technique similar to that used in Appendix to prove
(\ref{a8}).) This new variable transforms by an addition of a
constant with the gauge transformation of the potential: if 
$V^{{\rm out}}_a(s,l)\to V^{{\rm out}}_a(s,l)+\a(s,l)l_a$, then 
$\Phi_{V^{{\rm out}}}(l)\to \Phi_{V^{{\rm out}}}(l)+
\frac{1}{4\pi}\int\a(-\infty,l')\,\dl'$. The solution (\ref{FV}) is
the only one which satisfies (as shown by a simple calculation) 
\be
\int\frac{\Phi_{V^{{\rm out}}}(l)}{(v\s l)^2}\,\dl 
=\int\frac{v\s V^{{\rm out}}(-\infty,l)}{v\s l}\,\dl
\label{Ph}
\ee
for any velocity $v$. 

Next, we want to determine the outgoing Dirac field which may be
regarded as independent of $V^{{\rm out}}$ from the point of view of
Poincar\'e generators. To this end put 
$V(s,l)=V^{{\rm out}}(s,l)+V(+\infty,l)$ into (\ref{enm}) and 
(\ref{anm}). One finds $\frac{1}{2}\{ V,r_aV\} 
=\frac{1}{2}\{ V^{{\rm out}},r_aV^{{\rm out}}\}$, but 
$\frac{1}{2}\{ V,n_{ab}V\} 
=\frac{1}{2}\{ V^{{\rm out}},n_{ab}V^{{\rm out}}\} + 
\{ V^{{\rm out}},n_{ab}V(+\infty,.)\}$. Substituting (\ref{Gau}) for 
$V(+\infty,l)$ and using the identity 
$(n_{ab}V^e)_c(v,l)=(v_a\d_b-v_b\d_a)V^e_c(v,l)$ we obtain
$\{ V^{{\rm out}},n_{ab}V(+\infty,.)\}=
\int n(v)(v_ai\d_b-v_bi\d_a)i\{ V^e(v,.), V^{{\rm out}}\}\m(v)$. 
Finally, introducing a new variable 
\be
g(v)=e^{\txt i\{ V^e(v,.), V^{{\rm out}}\} }f(v)
\label{gf1}
\ee
we bring the generators to the form
\begin{eqnarray}
&&P_a=(g, p_ag) +{\txt\frac{1}{2}}\{ V^{{\rm out}},r_aV^{{\rm out}}\}\, ,
\label{enout}\\
&&M_{ab}=(g, m_{ab}g) +
{\txt\frac{1}{2}}\{ V^{{\rm out}},n_{ab}V^{{\rm out}}\}\, .
\label{anout}
\end{eqnarray}
Now, define the free outgoing Dirac field by 
\be
\p^{{\rm out}}(x) = \lp\frac{m}{2\pi}\rp^{3/2}
\int e^{\txt -imx\s v\g\s v}\g\s v g(v) \m(v)\, ,
\label{dirac}
\ee
which is a special, concise form of the Fourier representation and
which implies the asymptotic of the form (\ref{tas}) with $f(v)$
replaced by $g(v)$. Then the generators (\ref{enout}) and
(\ref{anout}) turn out to be the sums of the conserved quantities for
free fields $\p^{{\rm out}}(x)$ and $F^{{\rm out}}_{ab}(x)$.
Therefore, $\p^{{\rm out}}(x)$ should be interpreted as the field
describing free particles {\em together with their Coulomb fields}.
We have seen that the new separation of variables
(\ref{enout},\ref{anout}) forced the explicit appearence of a gauge
dependent quantity $\{ V^e(v,.), V^{{\rm out}}\}$, but only as a
phase transformation. With the use of (\ref{Ph}) the phase factor in
(\ref{gf1}) takes the form 
\be
e^{\txt i\{ V^e(v,.), V^{{\rm out}}\} }=\exp\lp\frac{ie}{4\pi}\int
\frac{\Phi_{V^{{\rm out}}}(l)}{(v\s l)^2}\, \dl\rp\, ,
\label{stp}
\ee
and this is the
only way in which a gauge-dependent quantity appears in the classical
asymptotic structure. It is natural and economic, therefore, to
assume, that the additive constant in $e\Phi_{V^{{\rm out}}}(l)$ is a
phase variable. Consequently, we put into one class gauges 
$V_1^{{\rm out}}$ and $V_2^{{\rm out}}$ such that 
$l\wedge V_1^{{\rm out}}(s,l) =l\wedge
V_2^{{\rm out}}(s,l)$ and $\Phi_{V_2^{{\rm out}}}(l) -
\Phi_{V_1^{{\rm out}}}(l) =n2\pi/e$, $n\in\Z$. 

With the above knowledge of the classical structure we can now return
to the problem of taking into account the Gauss constraint on the
quantum level. The form (\ref{Gau}) of this constraint is not suited
for the translation to an abstract algebraic level. However, the
physical interpretation prompts an indirect solution. Instead of
either the pair $(V,f)$ or $(V^{{\rm out}},g)$ it is natural to
work with the pair of variables having the direct physical meaning:
the asymptotic total electromagnetic field $V$ and the asymptotic
field of charged particles, with their Coulomb fields included, $g$.
The commutation relations (\ref{hq1}) and (\ref{hq2}) will be now
reformulated in terms of these variables with the use of the relation
\be
g(v)=e^{\txt i\{ V^e(v,.), V\} }f(v)\, ,
\label{gf2}
\ee
which is equivalent to (\ref{gf1}) by $\{ V^e(v,.),
V(+\infty,.)\}=0$. Two circumstances have to be taken into account.
First, on the classical level the Gauss constraint has been
completely solved, so for the electromagenetic test field asymptotic
the free field part only should be taken. Second, the quantization
has to be consistent with our identifying the gauges differing by
$n2\pi$ in $e\Phi(l)$. This problem is solved, as is easily seen from
(\ref{hq1}) and (\ref{hq2}), by using the electromagnetic variable in
the form  $e^{-i\{ V_1^{{\rm out}}, V^{{\rm q}}\} }$ only. Then, the
relations (\ref{hq1}) take the form 
\be
e^{\txt -i\{ V_1^{{\rm out}}, V^{{\rm q}}\} }
e^{\txt -i\{ V_2^{{\rm out}}, V^{{\rm q}}\} }= 
e^{\txt -i\frac{1}{2}\{ V_1^{{\rm out}}, V_2^{{\rm out}}\} }
e^{\txt -i\{ V_1^{{\rm out}}+V_2^{{\rm out}}, V^{{\rm q}}\} }\, ,
\label{hqw}
\ee
the relations (\ref{hq2}) remain true for $g^{{\rm q}}$, but now the
two variables do not commute:
\be
e^{\txt -i\{ V_1^{{\rm out}}, V^{{\rm q}}\} } g^{{\rm q}}(v)= 
e^{\txt +i\{ V_1^{{\rm out}}, V^e(v,.)\} }g^{{\rm q}}
e^{\txt -i\{ V_1^{{\rm out}}, V^{{\rm q}}\} }\, .
\label{hqi}
\ee
The last relation has an obvious physical interpretation: the element
$g^{{\rm q}}(v)$, beside its fermionic role, annihilates the Coulomb
field with the asymptotic $V^e(v,l)$. This is, clearly, the
implementation of Gauss' law on the quantum level. Note, also, that
by (\ref{stp}) the relation is indeed consistent with our
identification of gauge classes. Observe,
furthermore, that the element 
$e^{-i\{ V_1^{{\rm out}}, V^{{\rm q}}\} }$ creates the field with the
asymptotic $V_1^{{\rm out}}(s,l)$. The use of only free fields as
test fields reflects the fact, that the Coulomb field is fastened to
a particle, which is a neat confirmation of the consistency of our
scheme. Neverthelss, the element 
$e^{-i\{ V_1^{{\rm out}}, V^{{\rm q}}\} }$ is a functional of
the {\em total} field $V^{{\rm q}}$, as assumed in the construction.
This is seen from (\ref{hqi}), and also from 
\be
\{ V_1^{{\rm out}}, V\} =\{ V_1^{{\rm out}}, V^{{\rm out}}\} - 
\frac{1}{4\pi}\int V_1^{{\rm out}}(-\infty,l)\s V(+\infty,l)\, 
\dl\, .
\label{split}
\ee
It is clear from both arguments, that in order to ``catch'' the whole
field, it is absolutely necessary that all free test fields are
admitted, also those infrared-singular, for which $V^{{\rm
out}}(-\infty,l)\neq 0$. This fact is to become of crucial importance
for the interpretation of further results. 

The quantum structure thus obtained will be now given an 
unobjectionable algebraic form, formulated in terms of elements
heuristically identified by 
$$
W(V)= e^{\txt -i\{ V, V^{{\rm q}}\} }~~,~~~B(g)=(g,g^{{\rm q}})\, ,
$$
where from now on all the test fields $V(s,l)$ are free fields, so we
omit the superscript ``out''. We have to specify the scope of the
test functions. Let $\K$ be the Hilbert space of (equivalence classes
of) $\C^4$-valued functions $g(v)$ on the hyperboloid $v^2=1$,
$v^0>0$ with the scalar product (\ref{sp}). Let ${\cal V}$ be the
linear space of homogeneous of degree $-1$ (Eq.(\ref{hom})) functions
$V_a(s,l)$, infinitely differentiable in both variables outside $l=0$
(differentiations with respect to $l$ in the sense of the action of
the operator $L_{ab}$) and satisfying the conditions
\begin{eqnarray}
&&l\s V(s, l) = 0\, ,
\label{tran}\\
&&|L_{b_1c_1}\ldots L_{b_kc_k}\V_a(s, l)|
<\frac{\con (k)}{(t\s l)^2(1+|s|/t\s l)^{1+\e}}\, ,
~~k=0,1,\ldots\, ,~~
\label{dec}\\
&&V_a(+\infty,l)=0\, ,
\label{fr}\\
&&L_{[ab}V_{c]}(-\infty,l)=0\, ,
\label{eLf}
\end{eqnarray}
where the second condition holds for some ($V$- and $k$-dependent)
$\e>0$ and for an arbitrarily chosen unit timelike, future-pointing
vector $t$; the bounds are then true for any other such vector (with
some other constants $\con(k)$). These bounds guarantee the existence
of the limit functions as infinitely differentiable, homogeneous
functions of degree $-1$. Let $L$ be the Abelian additive group of
elements $(V)$ defined as pairs 
\be
(V)=\lp l\wedge V(s,l), \Phi_{V}(l)\bmod 2\pi/e\rp\, ,
\label{symg}
\ee
where $V\in{\cal V}$ and $\Phi_{V}$ is defined by (\ref{FV}). In
other words, $L$ is the quotient of the additive group ${\cal V}$
through the equivalence relation $\sim$, $L={\cal V}/\sim$, where 
\be
V_1\sim V_2~~~{\rm iff}~~~~
\begin{array}{@{}l}
\dsp V_2(s,l)-V_1(s,l)=l\,\a(s,l)~~{\rm and} \\
\dsp \frac{1}{4\pi}\int\a(-\infty,l)\,\dl =n\frac{2\pi}{e}\, .
\end{array}
\label{sim1}
\ee
The group $L$ inherits from ${\cal V}$ the
symplectic form (\ref{sympl}). Denote, also, for later use, 
${\cal V}_0\df\{ V\in {\cal V}| l\wedge V(-\infty,l)=0\}$ and 
$L_0\df {\cal V}_0/\sim$. 

The symplectic group $L$ and the Hilbert space $\K$ 
supply the test fields for the elements $W(V)$, $(V)\in L$ (the
parenthesis in the symbol $(V)$ appearing as an argument of $W$ or of
the symplectic form will be omitted) and $B(f)$, $f\in \K$ which
generate a particular *-algebra $\B$ of Sec.\ref{semi} according to
the relations 
\be
W(V)^* = W(-V)\, ,~W(0)=E\, ,~W(V_1)W(V_2)=
e^{\txt -\frac{i}{2}\{ V_1,V_2\}} W(V_1+V_2)\, ,
\label{weyl}
\ee
\be
[B(g_1), B(g_2)]_+ =0\, ,~~~[B(g_1), B(g_2)^*]_+ = (g_1, g_2) E\, ,
\label{ferm}
\ee
\be
W(V) B(g) = \b_{\Phi_V}\! \lp B(g)\rp W(V)\, ,
\label{com1}
\ee
where
\be
\b_\Phi\! \lp B(g)\rp = B(S_\Phi g)\, ,
~~\lp S_\Phi g\rp(v) = \exp\lp\dsp i\frac{e}{4\pi}
\int\frac{\Phi(l)}{(v\s l)^2}\,\dl\rp\, g(v)\, .
\label{com2}
\ee

\begin{defin}
The asymptotic field algebra is the $C^*$-algebra $(\F , \| .\| )$
obtained from the above *-algebra $\B$ according to Definition
\ref{field}.  
\label{asfi}
\end{defin}

Let us consider the role of elements $W((0, c \mo))$, which form an
Abelian one-parameter group $W((0, c_1 \mo))W((0, c_2 \mo))= 
W((0, c_1+c_2 \mo))$. The relations (\ref{ch}) and
(\ref{split}) suggest that in any representation in which
$\pi(W((0,c \mo)))$ are strongly continuous and written as 
$e^{\txt icQ_\pi}$, the operator $Q_\pi$ has the interpretation of
the charge operator. This interpretation is confirmed by the action
of the automophism $\g_c$ of $\F$, 
\be
A\rightarrow\g_c(A) \df W((0, c \mo))AW((0, c \mo))^*
\label{gatr}
\ee
on the basic elements:
$$
\g_c\lp W(V)\rp = W(V),~~ \g_c\lp B(g)\rp = e^{\txt -iec} B(g)\, .
$$
Now, as $2\pi/e =0 ({\rm mod}~2\pi/e)$, there is 
$e^{\txt i2\pi Q_\pi/e}=\I$, which implies that the spectrum of
charge is contained in the set $\{ne|n\in\Z\}$. 
The variable $e\Phi(l)$ is connected with the free
electromagnetic field, so bringing into play other carriers of
charge should respect the phase character of the additive constant in
this function. This means that the assumption in the following
corollary is well founded.
\begin{col}
If a $C^*$-algebra of asymptotic fields contains the subalgebra
generated by elements $W((0,c \mo))$, then the charge is quantized in
units of $e$.
\end{col}

In the following definition particular elements of $\F$ are
distinguished as observables in the obvious way.
\begin{defin}
The algebra of observables $\A$ is the $C^*$-subalgebra of $\F$ of
elements invariant under the gauge transformation (\ref{gatr}).
\label{obs}
\end{defin}

One has to stress at this point that all the Weyl elements are
therefore (functions of) observables. Denying the elements with
infrared-singular test functions ($V(-\infty,l)\neq 0$) the status of
observables one would deprive the total electromagnetic field of its
Coulomb part, as discussed earlier. We shall return to this important
point when discussing representations of our algebra. Also, all
elements $B(f)^*B(g)$ are in $\A$. 

The restricted Poincar\'e group (or rather its covering group) is
represented in the group of automorphisms of the field algebra $\F$.
One easily shows that the operators (\ref{poin1}) and (\ref{poin2}) 
(the variable $g(v)$ undergoes the same transformations as $f(v)$)
satisfy the consistency condition (\ref{bo3}), namely
$$
R_{x,A} S_\Phi = S_{T_{x,A}\Phi} R_{x,A}\, ,
$$
where the transformation $\lb T_{x,A}\Phi\rb(l)=\Phi(\Lambda^{-1}l)$
is implied by (\ref{poin1}) and (\ref{FV}). Therefore, the action of 
the Poincar\'e group on $\F$ may be consistently defined by 
$$
\a_{x,A}\lp W(V)\rp = W(T_{x,A}V)\, ,
~~\a_{x,A}\lp B(g)\rp = B(R_{x,A}g)\, .
$$

We end this section with the demonstration that for free test fields
the symplectic structure discussed above is an extension of the
structure used in more traditional algebraic formulations. To see
this we find the connection with the work of Roepstorff \c{roe70}.
This author uses the electromagnetic test fields of the form
$F_{ab}(x) = 4\pi\int D(x-y) (\n_a \ph_b(y) - \n_b \ph_a(y))d^4 y$,
where $D(x)$ is the Pauli-Jordan function and $\ph_a(x) =
\n^b\ph_{ab}(x)$ for some antisymmetric test function of compact
support $\ph_{ab}(x)$. With the use of representation $D(x) =
-\frac{1}{8\pi^2} \int\delta'(x\s l)\, \dl$ a Lorentz gauge potential
for this field takes the form (\ref{pot}), with
\be
V_a(s,l) = \int\delta(s-y\s l)\ph_a(y) d^4 y\, .
\label{roe}
\ee
Substituting two functions of this form in (\ref{sympl}) one has
\be
\{V_1, V_2\} = 4\pi\int\ph_1^a(x) D(x-y) \ph_{2a}(y) d^4 x d^4 y.
\label{sfr}
\ee
The r.h.side is the symplectic form used by Roepstorff (up to
multiplicative constants due to electromagnetic conventions).
However, the space of test fields is smaller in this formulation. It
is obvious from (\ref{roe}) that $V_a(-\infty, l) = 0$, so all these
fields are infrared-regular (the spacelike asymptotic of $F_{ab}$ has
no $1/r^2$ term). In fact, even a stronger regularity property holds.
There is $V_a(s,l) =\dot{J}_a (s,l)$, where $J_a(s,l)$ is a smooth
function vanishing outside a compact region (for a fixed scaling $t\s
l = 1$), given by 
$J_a(s,l) = \int\delta(s-y\s l) l^b \ph_{ab}(y) d^4 y$. 
It will prove convenient to reformulate the above formulas in the
Fourier-transformed version. It is shown in Sec.\ref{auw} below that
\be
\{ V_1, V_2\} = i\,\mbox{P}\int\ov{\tV_1}\s \tV_2(\w,l)
\frac{d\w}{\w}\, d^2 l\, ,
\label{symf}
\ee
where P denotes the principal value. If $\tV_a(\w,l)$ vanishes for
$\w\rightarrow 0$ sufficiently fast (e.g., as $|\w|^{\e}$), then
the principal value sign may be omitted. This is true, in
particular, for $V$ given by (\ref{roe}). In this case 
$\ti{V}_a(\w,l) =\hat{\ph}_a(\w l)$, where  
$\hat{\ph}(p) = \frac{1}{2\pi}\int \ph(x) e^{\txt ip\s x} d^4 x$, 
and then $\tV_a(\w,l) = -i\w\hat{\ph}_a(\w l)$. 
On the other hand, the last equation shows that our general field
satisfies the condition of Roepstorff's space $L_1$.

\setcounter{equation}{0}
\setcounter{pr}{0}
\setcounter{defin}{0}

\section{\hspace*{-.7cm}. Existence of charge and energy-momentum, 
and the regularity of representations} 
\label{grep}

In the present section we investigate the consequences of putting
some physical restrictions on representations. The following
definitions, the second of which is standard \c{haa}, will
simplify the formulation of propositions.
\begin{defin} \hspace*{\fill}\\
(i) A representation $\pi$ of $\F$ will be called a
charge-representation of $\F$ iff the one-parameter group 
$\R\ni c\rightarrow \pi\lp W((0,c\bmod 2\pi/e))\rp$ is strongly
continuous. \\ 
(ii) A representation $\pi$ of the algebra $\F$ acting in the
Hilbert space $\H$ is said to satisfy Borchers' criterion iff
there exists a unitary, strongly continuous representation
of the translation group $U(x)$ acting in $\H$, with the spectrum
contained in the closed forward lightcone, 
${\rm Spec}U(x)\subset \ov{V_+}$, and implementing translations of
all $A\in\F$:
\be
\pi(\a_x A) = U(x)\pi(A) U(-x)\, .
\label{phys0}
\ee
(iii) A representation $\pi$ of the algebra $\F$ 
will be said to satisfy Borchers' criterion
with respect to $\A$ (w.r.t. observables) iff the restiction of $\pi$
to the subalgebra $\A$ satisfies Borchers' criterion.
\label{phys}
\end{defin}
The defining properties of (i) and (iii) are necessary requirements
for a physically admissible representation. We analyze their
implications. 

\begin{pr}
If $\pi$ is a charge-representation of $\F$ 
then it has one of the following properties (or is
a direct sum of these three types):\\
(i) The charge takes on all the values $ke,\, k\in\Z$, and each
charge eigenspace is cyclic.\\
(ii) (resp.(iii)) The subspace of vectors satisfying 
$\pi\lp B(g)\rp\p = 0~~\forall g\in \K$ 
(resp. $\pi\lp B(g)^*\rp\p = 0~~\forall g\in\K$) is cyclic.
\label{charge}
\end{pr}
Proof. Given $\pi(\F)$ on $\H$ let $\H_k\subset\H$ be the subspace of
all charge eigenvectors to the eigenvalue 
$ke$ for a given $k\in\Z$, and let $\H'_k = [\pi(\F)\H_k]$ (the closed
linear subspace spanned by vectors in $\pi(\F)\H_k$). Then 
$\H = \H'_k\dis\H'^\bot_k$, where both subspaces are invariant.
Moreover, if we split $\H'^\bot_k = \H'_{k+}\dis\H'_{k-}$, where the
spectrum of charge goes from $(k+1)e$ upwards on $\H'_{k+}$ and from
$(k-1)e$ downwards on $\H'_{k-}$, then both subspaces are separately
invariant. This occurs because the generating elements
$\pi\lp W(V)\rp$, $\pi\lp B(g)\rp$ and $\pi\lp B(g)^*\rp$ carry
charge $0$ or $\pm e$, so they cannot match the gap between
$\H'_{k+}$ and $\H'_{k-}$.

The representations $\pi(\F)\Big|_{\H'_{k+}}$ and
$\pi(\F)\Big|_{\H'_{k-}}$ have the properties (ii) and (iii)
respectively. To see this, let $\H''_{k+}$ be the subspace of
$\H'_{k+}$ of vectors satisfying the equation in (ii). Let $\p$ be an
element of the invariant subspace 
$\H'_{k+}\cap [\pi(\F)\H''_{k+}]^\bot$, with a bounded spectral
content of charge (these vectors form a dense subspace as
$e^{icQ_\pi}\in\pi(\F)$). If $\p\neq 0$ then there exists such
$f\in\K$ that $\pi\lp B(f)\rp\p \neq 0$. Continuing in this way one
can lower the charge spectral content of the vector unlimitedly. 
This contradicts the charge content of $\H'_{k+}$. 
A similar proof holds for $\pi(\F)\Big|_{\H'_{k-}}$.

The rest of the proof is simple inductive reasoning. Let the set of
integers $\Z$ be organized into a sequence $\{k_n\}$. Suppose that
for $\pi(\F)$ on $\H(n)$ the charge eigensubspaces 
$\H_{k_1},\ldots ,\H_{k_n}$ are cyclic. Take the next charge
eigenspace $\H_{k_{n+1}}$ and decompose $\H(n)$ according to the
above prescription: $\H(n) = \H'_{k_{n+1}}  
\dis\H'_{k_{n+1}+}\dis\H'_{k_{n+1}-}$. The representations
$\pi(\F)\Big|_{\H'_{k_{n+1}\pm}}$ satisfy the properties (ii) and
(iii) respectively. By construction $\H_{k_{n+1}}$ is cyclic for
$\pi(\F)$ on $\H(n+1)\df\H'_{k_{n+1}}$. As each of the subspaces 
$\H'_{k_{n+1}}$, $\H'_{k_{n+1}+}$ and $\H'_{k_{n+1}-}$ is invariant,
in particular with respect to charge operator, so 
$\H_{k_i}=\lp \H_{k_i}\cap\H'_{k_{n+1}}\rp \dis 
\lp \H_{k_i}\cap\H'_{k_{n+1}+}\rp \dis 
\lp \H_{k_i}\cap\H'_{k_{n+1}-}\rp$, and the spaces
$\H_{k_i}\cap\H(n+1)$ are cyclic in $\H(n+1)$ for $i=1,\ldots ,n$. 
Continuing in this way we obtain a direct sum of  
representations satisfying the properties (ii) or (iii) and a
representation on $\dsp\H=\bigcap_{n=1}^\infty \H(n)$, which
satisfies the property (i) by construction. \ep

If, in addition to the charge, physical energy-momentum observables 
are assumed to exist, then one has the following result. 
\begin{pr}
The charge spectrum consists of all values $ke$, $k\in\Z$, and
eigenspace to each charge value is cyclic for a charge-representation
of $\F$ satisfying Borchers' criterion with respect to
observables. 
\label{phys1}
\end{pr}
The proof of the proposition will be based on the following 
observation.
\begin{lem}
If the representation $\pi$ of $\F$ has the property (ii)
of Proposition \ref{charge}, then it is unitarily equivalent to a
representation of type (\ref{repf}), where $\pi_f$ is the Fock
representation based on the cyclic vector $\W_f$ satisfying
$\pi_f\lp B(g)\rp\W_f = 0~~\forall g\in \K$.
\label{unphys}
\end{lem}
An analogous result holds for representations satisfying the property
(iii) of Prop.\ref{charge}. (In that case $\pi_f$ is the Fock
representation based on the cyclic vector satisfying $\pi_f\lp
B(g)^*\rp\W_f=0$ $\forall g\in \K$. We use the term Fock
representation in the wider sense, referring to any of the
representations differing by a Bogoliubov transformation from the one
appearing in the lemma.)\\
Proof. Let the representation $\pi$ act in $\H$ and denote the
respective cyclic subspace by $\H_r$. If $C\in\ca$ then it may be
represented as $C = \w_F(C)E+C'$, where $\w_F$ is the Fock state
and $C'$ is a sum of elements having $B(f)$ on the
right and/or $B(g)^*$ on the left. Therefore for
$\ph,\p\in\H_r$ there is $(\ph,\pi(C)\p) = \w_F(C)(\ph,\p)$.
$\H_r$ is invariant under $\pi\lp W(V)\rp$, hence the vectors
$\sum_{k=1}^N \pi(C_k)\ph_k$, $\ph_k\in\H_r$, 
are dense in $\H$. Let $\pi_r$ be the restriction of $\pi$ to the
Weyl algebra and to the space $\H_r$. We
map $\H$ onto $\H_F\dip\H_r$ ($\H_F$ is the Hilbert space of the
Fock representation $\pi_F$) by the rule
$\sum_k\pi(C_k)\ph_k\rightarrow\sum_k\pi_F(C_k)\W_F\dip\ph_k$.
It is now easy to show this is a unitary map providing the
claimed equivalence of representations.\ep

Proof of Prop.\ref{phys1}. In view of the result of 
Prop.\ref{charge} one has to show that a representation of the type
described in the lemma cannot be a subrepresentation of a
representation satisfying the assumptions of Prop.\ref{phys1}. 
By a general theorem by Borchers \c{bor66}
(see also \c{haa}) the
representation $U(x)$ may be chosen to lie in $\pi(\A)''$, hence
it suffices to show that for the representation of the lemma itself
there is no $U(x)$ in $\pi(\A)''$ implementing translations of
observables and satisfying the spectral condition. 
Suppose the converse is true. 
Then for any $C\in \mbox{CAR}\cap\A$ there is 
$U(x) \pi(C) U(-x)=\pi(\a_xC) =
\pi_F(\a_xC)\dip\I_r = [U_F(x)\dip\I_r][\pi_F(C)\dip\I_r]
[U_F(-x)\dip\I_r]$, where $U_F(x)$ is the unitary representation of
translations in the Fock representation. This means that
$U(x)\lb U_F(-x)\dip\I_r\rb\equiv R(x)$ commutes with
$\pi_F(C)\dip\I_r$ for all $C\in\mbox{CAR}\cap\A$. The representation
$\pi_F(\mbox{CAR}\cap\A)$ acts irreducibly on each of the subspaces 
$\H_n\subset \H_F$ spanned by all vectors $\pi_F(B(g_1)^*\ldots
B(g_n)^*)\W_F$, and all spaces $\H_n\dip\H_r$ are invariant with
respect to $R(x)$ (as they are invariant with respect to $\pi(\A)$). 
Hence, $U(x)\Big|_{\H_n\dip\H_r}=
U_F(x)\Big|_{\H_n}\dip U_{r,n}(x)$, where $U_{r,n}$ is a strongly
continuous representation of translations in $\H_r$. 
However, the Fock representation $\pi_F$ appearing in this
construction is not ``the right'' Fock representation of the free
Dirac field ($B(g)$ contains both positive and negative
frequencies), and the energy spectrum of $U_F(x)\Big|_{\H_n}$ 
is not bounded from below for $n\geq 1$, which contradicts 
the assumption.    \ep

The existence of a charge operator is a necessary, but by far not a
sufficient condition for the operators $\pi(W(V))$ to have a clear
physical interpretation. We recall that, as explained in the previous
section, in our algebra all these Weyl operators should be understood
as observables, more precisely, as exponentials of (unbounded)
observable electromagnetic field operators. The test function set $L$
is an Abelian group rather than a vector space, so a direct
multiplication of $V$ by a parameter is not possible. However, it is
sufficient to find a map $\R\ni\la\to (V^\la)\in L$, such that
$(V^0)=(0)$, $(V^1)=(V)$, $(V^{\la})+(V^{\nu})=(V^{\la+\nu})$ and 
$-(V^\la)=(V^{-\la})$. Then $W(V^\la)$ is a one-parameter group and
the above condition on the representation may be formulated as 
the strong continuity of $\pi(W(V^\la))$ in $\la$. Let 
$(V)=(l\wedge V, \Phi \mo)$. Then it is easily shown, that for each
choice of the representant $\Phi$ from the class $\Phi \mo$ the map 
\be
\R\ni\la\to (V^\la)_\Phi\df 
(l\wedge\lambda V, \lambda\Phi \mo )
\label{smult}
\ee
satisfies the listed requirements. 
\begin{defin}
Representation $\pi$ of the algebra $\F$ (or of a subalgebra of $\F$)
will be called regular iff all one-parameter groups 
$\R\ni\lambda\rightarrow \pi\lp W((V^{\lambda})_\Phi)\rp$ are strongly
continuous. 
\label{regr}
\end{defin}
Remarks. \hspace*{\fill}\\
(i) If $\pi$ is regular then it is a charge representation. This
follows by the special choice $(V)=(0, c \mo)$. \\
(ii) The generators of the groups $\pi\lp W((V^\la)_{\Phi'})\rp$ 
and $\pi\lp W((V^\la)_\Phi)\rp $, 
where $\Phi'\in\Phi\bmod 2\pi/e$, differ
by a multiple of the charge operator. This follows from 
$W((V^\la)_{\Phi'})= W((V^\la)_\Phi) 
W((0, \lambda 2k\pi/e\bmod 2\pi/e))$ for $\Phi'=\Phi +2k\pi/e$.

We are now prepared to partly characterize the representations
satisfying the condition of Definition \ref{phys}(ii). For the
positive energy Fock representation $\pi_F$ of the algebra CAR let us
denote  $\pi_F(B(g))= \int\ov{g(v)}\g\s v b(v)\m(v)$ and let 
$:\ldots:$ denote the standard normal ordering.
\begin{theor}\hspace*{\fill}\\
(i) A representation $\pi$ of the algebra $\F$ satisfies Borchers'
criterion if, and only if, it is unitarily equivalent to a
representation of type (\ref{repf}), where $\pi_f=\pi_F$ is the
positive energy Fock representation of \/{\rm CAR} and $\pi_r$ is a 
representation of the Weyl algebra satisfying Borchers' criterion.\\
For representations of this type the Gauss constraint in the
form (\ref{Gau}) is recovered in the von Neumann algebra $\pi(\A)''$:
\be
\pi(W(V))=\exp\lp -i\int :\ov{b(v)}\g\s v b(v):\{ V,V^e(v,.)\}\,
\m(v)\rp\dip\pi_r(W(V))\, .
\label{Gauq}
\ee
(ii) $\pi$ is irreducible iff $\pi_r$ is irreducible. \\
(iii) $\pi$ is regular iff $\pi_r$ is regular. 
\label{tcov}
\end{theor}
Remark. For $\pi(\F)$ in the form given by (i) the operator
$\I\dip\pi_r(W(V))$ has the interpretation of the free field
exponent, and one sees once more the essentiality of the regularity
assumption.\\
Proof. If the unitary equivalence is proved, then (\ref{Gauq})
follows by simple calculation. This formula then implies (ii) by
irreducibility of the Fock representation (and the identity \c{dix} 
$\lp {\cal L}(\H_F)\dip\pi_r({\rm CCR})\rp'=
\C\I_F\dip\pi_r({\rm CCR})'$) and (iii) by 
$\dsp\{ (V^\la)_\Phi,V^e(v,.)\} =-\la\frac{e}{4\pi}\int
\frac{\Phi(l)}{(v\s l)^2}\,\dl$. The equivalence of representations 
is proved by adapting to the present case the idea of
Ref.\c{bhs}. Let $g\in\K$. The positive and
negative frequency parts of $g$ are easily extracted from
$g$ by $g_\pm=P_\pm g$, where $P_\pm$ are projection operators
in $\K$ defined by $\lp P_\pm g\rp(v)=P_\pm(v)g(v)$,
$P_\pm(v)=\frac{1}{2}(1\pm\g\s v)$. Let $\pi$ satisfy Borchers'
criterion. From translational covariance one
shows in standard way that $\pi\lp B(g_+)\rp$ and $\pi\lp
B(g_-)\rp^*$ (resp. $\pi\lp B(g_+)\rp^*$ and $\pi\lp
B(g_-)\rp$) lower
(resp. raise) the energy content of a vector by at least $m$ (the
mass of the fermion). Let $\H_r$ be the subspace of the
representation space formed by all vectors $\ph$ satisfying 
$\pi\lp B(g_+)\rp\ph = \pi\lp B(g_-)\rp^*\ph=0~\forall g\in\K$. The
subspace $\H_r$ is invariant under $U$ and the 
subspaces $\lb\pi(\F)\H_r\rb$ and $\lb\pi(\F)\H_r\rb^\bot$ are 
invariant both under $\pi$ and $U$. Let $\p\neq 0$ lie in the dense
subspace of finite energy vectors in $\lb\pi(\F)\H_r\rb^\bot$. Then
for some $g\in\K$ there is $\pi\lp B(g_+)\rp\p\neq 0$ or $\pi\lp
B(g_-)\rp^*\p\neq 0$. By recurrence, this gives the way for reaching
negative energies, which contradicts the assumption. Hence $\H_r$ is
cyclic for $\pi(\F)$. The unitary equivalence to a representation of
type (\ref{repf}) follows now as in the proof of Lemma \ref{unphys}. 
However, the Fock representation appearing in the construction is in
the present case ``the right'', positive energy Fock representation.
The use of Proposition \ref{tu} finishes the proof. From this
proposition also the ``if'' statement of (i) follows easily. \ep

Our next objectives are the characterization of vacuum states on our
algebra, and the construction of a class of physically meaningfull
regular representations. The first problem is solved by the
adaptation to the present situation of standard analytical methods. 
We shall have to discuss some properties of the symplectic space of
test functions. For $V\in {\cal V}$ let us denote by $[\V]$ the
respective {\em field} test function, i.e. the equivalence class of
$\V$ with respect to the equivalence 
\be
\V'\approx \V~~{\rm iff}~~l\wedge\V'=l\wedge\V\, .
\label{sim2}
\ee
The vector space of these classes will be denoted by $\Lf$, and its
subspace of classes $[\V]$ with $l\wedge V(-\infty,l)=0$ by $\Lf_0$.  
The space $\Lf$ inherits the symplectic form from ${\cal V}$, 
$\{ [\V_1],[\V_2]\} = \{ V_1,V_2\}$. Now, the analytical properties
of this space needed for the characterization of vacuum states will
also play role for its extension $\La$ to be used in the construction
of regular representations. Therefore, in the next section we
introduce this auxiliary symplectic space and formulate the necessary
properties in this wider setting. 

\setcounter{equation}{0}
\setcounter{pr}{0}
\setcounter{defin}{0}

\section{\hspace*{-.7cm}. Extension and analytical properties 
of symplectic structure}
\label{auw}

The first step towards extension of the symplectic space of test
fields $\Lf$ will be generalizing the fall-off condition
(\ref{dec}) for $k=0$. We do it first for scalar
functions. Consider the real Hilbert space $L^2_{\e, t}$ of
(equivalence classes of) real, measurable functions $f(s, l)$,
homogeneous of degree $-2$, with finite norm
\be
\| f\|^2_{\e, t}= \int f^2(\tau t\s l, l) (|\tau|+1)^{1+\e} (t\s l)^2
\, d\tau\, \dl\, ,
\label{norm1}
\ee
where $\e>0$ and $t$ is a unit future-pointing vector. If $\ti{t}$ is
another future-pointing vector and $c_{\ti{t}\cdot t}\equiv \ti{t}\s
t+ \sqrt{(\ti{t}\s t)^2 -1}$ then for every null vector $l$ there is 
$c_{\ti{t}\cdot t}^{-1}\leq\ti{t}\s l/t\s l\leq c_{\ti{t}\cdot t}$. 
Using these bounds one shows that 
$\dsp c_{\ti{t}\cdot t}^{-2-\e} \| f\|^2_{\e, t}\leq 
\| f\|^2_{\e, \ti{t}} \leq c_{\ti{t}\cdot t}^{2+\e} \| f\|^2_{\e, t}$.
Therefore $L^2_{\e, t}$ does not depend on $t$ when considered as a
linear topological space. Components of smooth functions satisfying
(\ref{dec}) obviously lie in this space. 

If $f\in L^2_{\e, t}$ then for $\delta<\e$ 
\begin{eqnarray*}
&&\int |f(\tau t\s l, l)| (|\tau|+1)^{\frac{\delta}{2}} 
 |f(s t\s l, l)|
(|s|+1)^{\frac{\delta}{2}} (t\s l)^2\, d\tau\, ds\,
\dl \\
&&\leq\lp\int\lb\int|f(\tau t\s l, l)|^2(t\s l)^2\,
\dl\rb^{\frac{1}{2}} (|\tau|+1)^{\frac{\delta}{2}}\, d\tau\rp^2 \leq
\frac{2}{\e-\delta}\| f\|^2_{\e, t}\, .
\end{eqnarray*}
Consequently, the integral 
$\dsp\int |f(\tau t\s l, l)| (|\tau|+1)^{\frac{\delta}{2}}\, d\tau\,
t\s l$ is finite $\dl$-almost everywhere and as a function of $l$
(homogeneous of degree $-1$) is square integrable with respect to
$\dl$. In particular, the Fourier transform of $f(s, l)$ with respect
to $s$ is defined by the integral (\ref{four2}), satisfies
\be
\int\left|\ti{f}\lp\frac{\w}{t\s l}, l\rp\right|^2\, \dl 
\leq\frac{1}{2\pi^2\e}\|f\|^2_{\e, t}\, ,
\label{norm2}
\ee
and 
\be
\int\left|\ti{f}\lp\frac{\w}{t\s l}, l\rp\right|^2\, d\w = 
\frac{1}{2\pi} \int f^2(\tau t\s l, l)\, d\tau (t\s l)^2\, .
\label{norm3}
\ee
The function $\ti{f}(\w/t\s l,l)$ depends
continuously on $\w$, both pointwise in $l$ and as an element of the
Hilbert space $L^2(\dl)$. In fact, even stronger conditions are
satisfied. With the use of the bound $\dsp|e^{\txt ix}-1|<2|x|^\a$
for $0\leq\a<1$, one finds for any $\delta \in \langle 0, \min\{ 2,
\e\})$
\be
\left|\ti{f}\lp\frac{\w'}{t\s l}, l\rp-
\ti{f}\lp\frac{\w}{t\s l}, l\rp\right| 
\leq\frac{1}{\pi}|\w'-\w|^{\frac{\delta}{2}} \int 
|\tau|^{\frac{\delta}{2}}|f(\tau t\s l,l)|\, d\tau\, t\s l\, ,
\label{norm4}
\ee
and then
\be
\int \left|\ti{f}\lp\frac{\w'}{t\s l}, l\rp
-\ti{f}\lp\frac{\w}{t\s l}, l\rp\right|^2\,
\dl \leq\frac{2}{(\e-\delta)\pi^2}\| f\|^2_{\e, t}\,
|\w'-\w|^\delta\, . 
\label{norm5}
\ee
This implies, with the use of (\ref{norm2}), 
\be
\left| \int \left|\ti{f}\lp\frac{\w'}{t\s l}, l\rp\right|^2\,\dl - 
\int \left|\ti{f}\lp\frac{\w}{t\s l}, l\rp\right|^2\,\dl \right | 
\leq \frac{2}{\pi^2\sqrt{\e(\e-\delta)}} \| f\|^2_{\e, t} \,
|\w'-\w|^{\frac{\delta}{2}}\, .
\label{norm6}
\ee

Next, we derive some integral identities. 
Let $\ti{K}_{\a, \b}(\w)=\w^{-1}\chi_{\a, \b}(\w)$, 
where $\chi_{\a, \b}$ is the characteristic function of the set 
$\langle -\b, -\a\rangle\cup \langle \a, \b\rangle$, and let 
$f,\, g\in L^2_{\e,t}$. For almost all $l$ the function 
$\frac{1}{(2\pi)^2} f(\tau, l) \ti{K}_{\a, \b}(\w)
e^{\txt -i\w (\tau -s)} g(s, l)$ is
absolutely integrable with respect to $d\tau\, ds\, d\w$, so the
iterated integrals are equal, i.e.,
\be
\frac{1}{(2\pi)^2}\int f(\tau, l) K_{\a, \b}(\tau-s) g(s, l)\,
d\tau\, ds = \int_{|\w|\in\langle\a, \b\rangle} \ov{\ti{f}(\w, l)} 
\ti{g}(\w, l)\frac{d\w}{\w}\, .
\label{K}
\ee
$K_{\a, \b}(s)$ is uniformly bounded and $\dsp\lim_{\a\to 0} 
\lim_{\b\to \infty} K_{\a, \b}(s) = -i\pi\e(s)$, so
\be
\frac{1}{4\pi}\int f(\tau, l)\e(\tau-s) g(s, l)\, d\tau\, ds = 
i\, \mbox{P}\int\ov{\ti{f}(\w, l)} \ti{g}(\w, l)\frac{d\w}{\w}\, ,
\label{P1}
\ee
where P denotes the principal value operation. Integrating over $\dl$
we obtain
\be
\frac{1}{4\pi}\int \dl\,\int f(\tau, l)\e(\tau-s) g(s, l)\, 
d\tau\, ds = i\int\dl\,\mbox{P}\int\ov{\ti{f}(\w, l)} 
\ti{g}(\w, l)\frac{d\w}{\w}\, .
\label{P2}
\ee
Another form of the Fourier representation of this integral will be
usefull. We integrate $\dsp f(\tau' t\s l, l) \ti{K}_{\a, \b}(\w)
e^{\txt -i\w (\tau'-s')}  g(s' t\s l, l)(t\s l)^2$ over 
$ds'\, d\tau'\, \dl\, d\w$,\\
and then calculate the limits (in the same order as before). Changing
the variables $\tau', s'$ back to $\tau, s$ we have
\be
\frac{1}{4\pi}\int \dl\,\int f(\tau, l)\e(\tau-s) g(s, l)\, 
d\tau\, ds = i\,\mbox{P}\int\frac{d\w'}{\w'} 
\int\ov{\ti{f}\lp\frac{\w'}{t\s l}, l\rp} 
\ti{g}\lp\frac{\w'}{t\s l}, l\rp\,\dl\, .
\label{P3}
\ee

Consider now the linear space of vector functions $u_a(\tau,l)$ such
that $l\s u(\tau,l)=0$ and each component of 
$u_a$ (in any Minkowski frame) is an element of
the space $L^2_{\e, t}$. Divide this space into classes with
respect to the equivalence relation (\ref{sim2}): 
$[u_a]= u_a(\tau,l)~\mbox{mod}~\a(s,l)l_a$. We denote this factor
space by $\T_\e$. For decreasing $\e$ the spaces $\T_\e$ form an
increasing family of vector spaces. Their union
$\dsp\T\df\bigcup_{\e>0}\T_\e$ is therefore again a vector space. We
now list some properties of $\T$ and of some structures on it. Simple
proofs based on the preceding discussion are omitted. 

$\T$ becomes a symplectic space with the form 
\be
\begin{array}{@{}l}
\dsp\{[u_1], [u_2]\} = \frac{1}{4\pi}\int\dl\int u_{1a}(\tau,l)
\e(\tau-s)u_2^a(s,l)\, d\tau\, ds\\
\dsp ~~~~~~~~~~~~~~=-i\,\mbox{P}\int\frac{d\w'}{\w'}\int \lb -
\ov{\ti{u}_{1a}\lp\frac{\w'}{t\s l},l\rp } 
\ti{u}_2^a\lp\frac{\w'}{t\s l},l\rp \rb\, \dl\, .
\end{array}
\label{gen2}
\ee
The form is nondegenerate on $\T$. Consider, further, the linear
space of real measurable vector functions $f_a(l)$ on the cone,
homogeneous of degree $-1$, orthogonal to $l$, $l\s f(l)=0$, and such
that each component is an element of the Hilbert space $L^2(\dl)$.
Divide this space into equivalence classes 
$[f_a](l)\df f_a(l)~\mbox{mod}~\b(l)l_a$.  
This factor space is a Hilbert space, denoted by $\H_0$, with the
scalar product 
\be
\lp[f_1], [f_2]\rp_0 =\int\lb -f_{1a}(l)f_2^a(l)\rb\,\dl\, .
\label{gen3}
\ee
Now, one easily shows with the use of (\ref{norm2}) that if
$[u]\in\T$, then $[\ti{u}(0,.)]\in\H_0$. The map 
\be
\T\ni [u]\rightarrow p([u])=[\ti{u}(0,.)]\in\H_0 
\label{gen4}
\ee
is linear and onto, $p(\T)=\H_0$ (for a given $[f]\in\H_0$ put
$u_a(s,l)= \\
f_a(l) (t\s l)^{-1} h(s/t\s l)$, with $h$ of compact
support and $\ti{h}(0)=1$; then \\
$p([u])=[f]$). If at least one of the pair of
functions $[u_i]\in\T$, $i=1,2$, satisfies $p([u_i])=0$, then, 
by (\ref{norm4}), the following integral is well defined
\be
F\lp [u_1], [u_2]\rp =\int_{\w\geq 0}\lb -
\ov{\ti{u}_{1a}(\w,l)} \ti{u}_2^a(\w,l)\rb\frac{d\w}{\w}\, \dl\, ,
\label{F1}
\ee
and the symplectic form may be expressed by
\be
\{[u_1], [u_2]\} = 2\,\mbox{Im}F\,\lp [u_1], [u_2]\rp\, ,
\label{F2}
\ee
where the hermiticity of $F$ has been used:
\be
\ov{F\lp [u_1], [u_2]\rp} = F\lp [u_2], [u_1]\rp\, .
\label{F3}
\ee

For almost all $l$ the integral 
$\dsp U_a(s,l)=-\int_s^{+\infty} u_a(\tau,l)\, d\tau$ is well
defined for all $s$, and for almost all $s$ there is 
$\frac{\D}{\D s} U_a(s,l)
=u_a(s,l)$. It is now easy to see that the functions $U_a(s,l)$
generalize the test functions $V_a(s,l)$, and the symplectic form
(\ref{gen2}) extends the form (\ref{sympl}). The limit values at
$-\infty$ may be expressed by $U_a(-\infty,l)=-2\pi
\ti{u}_a(0,l)$. The only essential property of $V_a$ which has not
been taken into account yet is (\ref{elfi}), 
$l_{[a}\tV_{b]}(0,l)=-2\pi l_{[a}\D_{b]}\Phi(l)$. In order to
formulate its generalization for $U_a$ consider the Hilbert subspace
$\HI$ of $\H_0$ obtained as the closure of the linear subspace of
elements $[\D_a\phi]$, where $\phi$ is a real smooth homogeneous
function of degree $0$ (different extensions of $\phi(l)$ outside the
cone yield $\D_a\phi(l)$ on the cone in one class $[\D_a\phi]$), 
\be
\HI\df\ov{\{[\D_a\phi]\in\H_0| \phi\in C^\infty\} }^{\H_0}\, .
\label{IR}
\ee
For $\phi\in\Ci$ the class $[\D_a\phi]$ determines $\phi$ uniquely up
to a constant. With the notation 
$[\phi]\equiv \phi~\mbox{mod const}$, 
the map $[\D_a\phi]\rightarrow [\phi]$ is linear and injective. We
show in the Appendix that this map extends canonically to an 
injective map $j$ of $\HI$ into the space of classes $[\phi]$ with
$\phi$ square-integrable with respect to any (and all) of the
measures $\dl/(t\s l)^2$. In this way every element of $\HI$
corresponds uniquely to some $[\phi]\in j(\HI)$, and may be written
in this generalized sense as $[\D_a\phi]$. 

The extended symplectic space is now chosen as the subspace of $\T$
given by $\La\df p^{-1}(\HI)$, with $p$ defined in (\ref{gen4}),
and equipped with the symplectic form (\ref{gen2}). The elements of
$\La$ satisfying $p([u])=0$ form a linear subspace, denoted by
$\La_0$. These structures extend the field test functions spaces 
$L^f\subset\La$ and $L^f_0=L^f\cap\La_0$. 

The following properties of the form $F(.,.)$ will be needed for the
characterization of vacuum states. 
\begin{lem} \hspace*{\fill}\\
(i) For each pair $[u_i]\in \La$, $i=1,2$, the function 
$x\rightarrow F([u_1], [T_xu_2-u_2])$ is the boundary value for
${\rm Im} z=0$ of the function 
$$
\R^4+i\ov{V_+}\ni z\rightarrow F([u_1],[T_zu_2-u_2])\df 
-\int_{\w\geq 0}(e^{\txt i\w l\s z}-1)\ov{\tu_1}\s
\tu_2(\w,l)\frac{d\w}{\w}\, \dl\, ,
$$
which is continuous on its domain and analytic on $\R^4+iV_+$. \\
The following bounds (ii)--(iv) hold on the whole domain 
$z=x+iy\in\R^4+i\ov{V_+}$.\\
(ii) For any fixed, unit, future-pointing vector $t$
\begin{eqnarray*}
&&\left|e^{\txt F([u_1], [T_zu_2-u_2])}\right|\leq 
C(t, [u_1], [u_2])\times\\
&&\times \lb\lp 1+y^0+|\vec{y}|\rp^2 +\lp|x^0|+ |\vec{x}|
\rp^2\rb^{\txt \frac{1}{2}\int\tu_1\s \tu_2(0,l)\, 
\theta\lp \tu_1\s \tu_2(0,l) \rp\, \dl}\, ,
\end{eqnarray*}
where $y^0\equiv y\s t$, $|\vec{y}|\equiv\sqrt{(y\s t)^2-y^2}$ 
(and the same for $x$), and $\theta(.)$ is the Heaviside step
function. \\ 
(iii) If $\tu_1\s \tu_2(0,l)\leq 0$ $\dl$-almost everywhere, then 
\begin{eqnarray*}
&&\left|e^{\txt F([u_1], [T_zu_2-u_2])}\right|
\leq C(t, [u_1], [u_2])\times\\
&&\times \lb\lp 1+y^0-|\vec{y}|\rp^2 +\theta(x\s x)\lp|x^0|- |\vec{x}|
\rp^2\rb^{\txt -\frac{1}{2}\int|\tu_1\s \tu_2(0,l)| 
\, \dl}\, .
\end{eqnarray*}
(iv) If $\tu_1\s \tu_2(0,l)\geq 0$ $\dl$-almost everywhere, then 
$$
\left|e^{\txt -F([u_1], [T_zu_2-u_2])}\right|
\leq C(t, [u_1], [u_2])\, .
$$
(v) If $p([u_1+u_2])=0$, then for $x\in\R^4$ 
\begin{eqnarray*}
&&F([u_1+T_xu_2], [u_1+T_xu_2]) = F([u_1], [T_xu_2-u_2])\\ 
&&+ F([T_xu_2-u_2], [u_1]) + F([u_1+u_2], [u_1+u_2])\, .
\end{eqnarray*}
(vi) The Fourier representation (\ref{F1}) of $F(.,.)$ is the usual\/
one-photon scalar product when restricted to $\La_0$, which 
yields a dense subspace of the one-photon Hilbert space, and 
$$
F([T_xu_1], [T_xu_2])= F([u_1], [u_2])\, .
$$
\label{aux6}
\end{lem}
Proof. If $z\in\R^4+iV_+$, then for any $k\in\R^4$ the function 
$\C\ni\xi\rightarrow F([u_1], [T_{z+\xi k}u_2-u_2])$ is analytic in
some neighbourhood of $\xi=0$, which implies (i) by Hartog's theorem.
Properties (ii)--(iv) follow easily from the bound
$$
\left| F([u_1], [T_zu_2-u_2])-\int \tu_1\s \tu_2(0,l)\, 
\mbox{ln}\left|1-i\frac{z\s l}{t\s l}\right|\, \dl\right| 
\leq \mbox{const}(t, [u_1], [u_2])\, ,
$$
and the inequalities $1\leq \lp 1+y^0-|\vec{y}|\rp^2 +\theta(x\s x)
\lp|x^0|-|\vec{x}| \rp^2 \leq \left|1-iz\s l/t\s l\right|^2 
\leq \lp 1+y^0+|\vec{y}|\rp^2 +\lp|x^0|+ |\vec{x}|\rp^2$. 
To prove the bound we split $F$ into two parts
\begin{eqnarray*}
&&F([u_1], [T_zu_2-u_2]) \\
&&= -\int_{\w\geq 0}(e^{\txt i\w l\s z}-1)\lb \ov{\tu_1}\s
\tu_2(\w,l)- \tu_1\s \tu_2(0,l)e^{\txt -\w t\s l}\rb\,
\frac{d\w}{\w}\, \dl\\ 
&&-\int\lb \tu_1\s \tu_2(0,l)\int_0^\infty 
\lp\exp\lp i\w'\frac{z\s l}{t\s l}\rp -1\rp e^{\txt -\w'}\, 
\frac{d\w'}{\w'}\rb\, \dl\, .
\end{eqnarray*}
The first term is absolutely bounded by\\
$\dsp 2\int\left|\ov{\tu_1}\s \tu_2(\w,l)- 
\tu_1\s \tu_2(0,l)e^{\txt -\w t\s l}\right|\frac{d\w}{\w}\, \dl 
<\infty$. 
The second term is equal to $\int \tu_1\s \tu_2(0,l)\, 
\mbox{ln}\lp 1-iz\s l/t\s l\rp \, \dl$. The imaginary part of
$\mbox{ln}\lp 1-iz\s l/t\s l\rp$ yields a term bounded in
$z$, which ends the proof of the bound. Property (v) is
easily proved by straightforward calculation in the special case 
$u_1 =-u_2$, and then the general case follows from $u_1+T_xu_2 =
(u_1+u_2)+(T_x u_2-u_2)$. Statement (vi) follows from our
discussion of the relation of the present formulation with the
traditional one, ending Sec.\ref{asym}. \ep

\setcounter{equation}{0}
\setcounter{pr}{0}
\setcounter{defin}{0}

\section{\hspace*{-.7cm}. Vacuum versus regular, 
positive energy representations}
\label{vreg}

Now we can take up the study of physical representations of the
asymptotic algebra. First of all, vacuum states have to be
characterized. We denote by $\F_0$ the subalgebra of $\F$ generated
by CAR and by elements $W(V)$ with $(V)\in L_0$. 
\begin{theor}
(i) If a cyclic representation $\pi$ of the algebra $\F$ satisfies
Borchers' criterion with respect to the Weyl algebra of the
electromagnetic field, and $U(x)\W=\W$ for some choice of the
pertinent representation of translations $U(x)$ and of the cyclic
vector $\W$, then for each $C\in{\rm CAR}$  
\be
\lp\W , \pi(C W(V))\W\rp =0~~\mbox{if}~~
l\wedge V(-\infty, l)\neq 0\, .
\label{vaci}
\ee
(ii) A cyclic representation $\pi$ of the algebra $\F$ satisfies
Borchers' criterion and $U(x)\W=\W$ for some choice of the pertinent
representation of translations $U(x)$ and of the cyclic vector $\W$
if, and only if, there is $\lp\W , \pi(C W(V))\W\rp=w_F(C)w_r(W(V))$,
where $\w_F$ is the positive energy Fock state on ${\rm CAR}$ and
$\w_r$ is the state on ${\rm CCR}$ given by 
\be
\w_r(W(V)) =\left\{
\begin{array}{ll}
\dsp 0\, , &\mbox{if}~~l\wedge V(-\infty, l)\neq 0\, ,\\
\dsp f(V)  e^{\txt -\frac{1}{2} F([\V], [\V])}\, , 
&\mbox{if}~~l\wedge V(-\infty, l)=0\, ,
\end{array}
\right.
\label{vacii}
\ee
where $f:L_0\rightarrow\C$ is a function of positive type,
satisfying the condition
\be
f(V_1+(T_x -1)V_2)=f(V_1)~~~~\forall\, (V_1)\in L_0,\, 
(V_2)\in L,\, x\in M\, .
\label{vacp}
\ee
\label{vac1}
\end{theor}
Proof. (i) Using the algebraic relations (\ref{weyl}) and 
(\ref{com1}), and the relation (\ref{F2}) one finds 
\begin{eqnarray*}
&&e^{\txt F([\V_1],[T_x\V_2-\V_2])}CW(V_1)W(T_xV_2)\\
&&=e^{\txt -i\{ V_1,V_2\} }\ov{e^{\txt F([\V_1],[T_x\V_2-\V_2])}}
W(T_xV_2)\b_{-\Phi_2}(C)W(V_1)\, .
\end{eqnarray*}
It follows now from the invariance of $\W$ under $U(x)$ and from the
spectral properties of $U(x)$ that the value of
$\w(.)\df\lp\W,\pi(.)\W\rp$ on the l.h.side of this
identity is the boundary value for $\mbox{Im}z=0$ of the function 
$$
\R^4+i\ov{V_+}\ni z\rightarrow e^{\txt F([\V_1], [T_z\V_2-\V_2])} 
\lp\W, \pi(CW(V_1)) U(z) \pi(W(V_2))\W\rp\, ,
$$
continuous on its domain and analytic inside. By Lemma \ref{aux6}(ii)
this function is polynomially bounded. The expectation value of the
r.h. side of the above identity has similar properties in the
time-reflected region $\R^4-i\ov{V_+}$. By the edge of the wedge
theorem there is an open region containing $\R^4+i\ov{V_+\cup V_-}$
and a function analytic in this region which is the analytic
continuation of both these functions. This function is polynomially
bounded on $\R^4+i\ov{V_+\cup V_-}$, hence it is a polynomial. To
prove this implication let $h(z)$ be a function with the stated
properties, and choose a basis of Minkowski vector space consisting
of four unit future-pointing vectors $t_i$, $i=1,\ldots,4$. For fixed
real $\a^2$, $\a^3$, $\a^4$ the function 
$\C\ni\xi\rightarrow h(\xi t_1+\a^2t_2+\a^3t_3+\a^4t_4)$ is analytic
and polynomially bounded, hence by Cauchy inequality for analytic
functions it is a
polynomial in $\xi$. Therefore, by repeated use of the implication,
$h(x)$ is a polynomial, hence $h(z)$ is the same polynomial for
complex $z$. We have thus proved that 
\be
e^{\txt F([\V_1],[T_x\V_2-\V_2])} \w\lp CW(V_1) W(T_xV_2)\rp=
P_{C,V_1,V_2}(x)\, ,
\label{va2}
\ee
where $P_{C,V_1,V_2}$ is a polynomial. If $p([\V_1])=p([\V_2])\neq 0$, 
then by Lemma \ref{aux6}(iii) $P_{C,V_1,V_2}(x)=0$ (take
$|\vec{x}|=\con$ and $|x^0|\to +\infty$). For 
$x=0$, $[\V_1]=[\V_2]=\frac{1}{2}[\V]$, $p([\V])\neq 0$ we get 
$\w(CW(V))=0$, which ends the proof if (i). \\
(ii) If $U(x)$ implement translations and $U(x)\W=\W$ for cyclic
$\W$, then writing each $C\in\ca$ in the form $C=\w_F(C)E+C'$, where
$\w_F$ is the positive energy Fock state and $C'$ is a sum of
elements having $B(g_+)$, $B(g_-)^*$ on the right and/or 
$B(g_+)^*$, $B(g_-)$ on the left, one finds that 
$\w(CW(V))=\w_F(C)\w(W(V))$. By (\ref{weyl}), (\ref{F2}), 
and Lemma \ref{aux6}(v), for $p([\V_1+\V_2])=0$ Eq.(\ref{va2}) is now
equivalent to   
\be
f(V_1+T_xV_2)= P'_{V_1,V_2}(x)\, ,
\label{va3}
\ee
where $P'_{V_1,V_2}$ is a polynomial and $f$ is a function on 
$L_0$ defined by 
\be
f(V)=\w(W(V)) e^{\txt \frac{1}{2} F([\V], [\V])}\, .
\label{va4}
\ee
By Lemma \ref{aux6}(vi) $f(V_1+T_xV_2)$ is absolutely bounded as a
function of $x$ for fixed $(V_1), (V_2)\in L_0$. Hence, by
(\ref{va3}),    
\be
f(V_1+T_xV_2)=f(V_1+V_2)~{\rm for}~(V_1),(V_2)\in L_0\, .
\label{va5}
\ee
The positivity of $\w(A^*A)$ for $A=\sum_{i=1}^n \a_iW(V_i)$,
for $(V_i)\in L_0$ and all sequences of complex
numbers $\{\a_i\}_{i=1}^n$, is equivalent to the condition \\
$\sum_{i,k=1}^n \ov{\b_i}\b_k e^{\txt F([\V_i], [\V_k])} 
f(V_k-V_i)\geq 0$ for all sequences of complex numbers 
$\{\b_i\}_{i=1}^n$. Following \c{rst} 
we replace all $V_i$ in this condition by 
$\frac{1}{N}\sum_{k=0}^{N-1} T_x^kV_i$ and take the limit 
$N\to\infty$. Then by (\ref{va5}), Lemma \ref{aux6}(vi), and the
ergodic theorem one has $\sum\ov{\b_i}\b_k f(V_k-V_i)\geq
0$, hence $f$ is of positive type (the remaining conditions,
$f(0)=1$ and $\ov{f(V)}=f(-V)$, are obviously satisfied). 
Consequently, $f$ is bounded and, by (\ref{va3}), (\ref{va5}) is
satisfied for all $(V_1),(V_2)\in L$ with
$(V_1+V_2)\in L_0$. This is equivalent to (\ref{vacp}),
which ends the proof that the state $\w$ has the form
given in (ii). 

Conversely, let $\w_r$ be a linear functional of the form given by
(ii). Then it is a state (one uses the fact that if $A$ and $B$ are
two Hermitian positive matrices $n\times n$, then the matrix $C$
defined by $C_{ik}= A_{ik}B_{ik}$ is also positive). Translations are
implemented in $\pi_r$ by the canonically defined representation of
translations $U_r(x)$ ($\w_r$ is translationally invariant). 
This representation is strongly continuous and satisfies the spectrum
condition. To show this, it is sufficient to demonstrate that for any
pair  $(V_1), (V_2)\in L$ the function $x\rightarrow
\w_r(W(V_1)W(T_xV_2))$ is continuous, and its Fourier transform has
support in $\ov{V_+}$. If $(V_1+V_2)\notin L_0$, then
this function vanishes identically. For $(V_1+V_2)\in L_0$ 
we have by (\ref{F2}) and Lemma \ref{aux6}(v)  
\begin{eqnarray*}
&\w_r(W(V_1)W(T_xV_2))&=e^{\txt -\frac{i}{2}\{ V_1, V_2\} } 
e^{\txt -\frac{i}{2}\{ V_1, T_xV_2-V_2\} }\w_r(W(V_1+T_xV_2))\\
&&=\w_r(W(V_1)W(V_2)) e^{\txt -F([\V_1],[T_x\V_2-\V_2])}\, .
\end{eqnarray*}
The function $\R^4+i\ov{V_+}\ni z\rightarrow
e^{\txt -F([u_1],[T_zu_2-u_2])}$ is analytic and, by
Lemma~\ref{aux6}(iv), bounded on its domain. Therefore it is the
Laplace transform of a  distribution with support in $\ov{V_+}$
\c{ree}. Now, the Fock state $\w_F$ on ${\rm CAR}$ 
satisfies (\ref{cf}), so the state on $\F$ defined by
$\w(CW(V))=\w_F(C)\w_r(W(V))$ generates a representation unitarily
equivalent to a representation of the type (\ref{repf}). Translations
are implemented in this representation by $U_F(x)\dip U_r(x)$, where
$U_F(x)$ is the representation canonically corelated to $\w_F$, which
ends the proof. \ep 

The representation space $\H$ of a vacuum representation, as implied
by the general result (i) of the above theorem, is easily seen to be
the uncountable direct sum of the space $[\pi(\F_0)\W]$ and spaces
derived from it by the action of operators $\pi(W(V))$; if 
$l\wedge V_1(-\infty,l)= l\wedge V_2(-\infty,l)$ the respective
spaces are equal, in other case they are orthogonal. In consequence,
vacuum representations are nonregular with respect to the
Weyl operators with infrared-singular test functions 
($l\wedge V(-\infty,l)\neq 0$). Now, the derivation of our 
asymptotic algebra has led unambiguously to its interpretation, as
explained in Sec.\ref{asym} and \ref{grep}. From the point of view of
this interpretation the above structure does not seem a physically
justified idealization, for two reasons:\\
(i) The infrared singular Weyl operators, being degraded to
operators intertwining between different representations of $\F_0$, 
no longer describe the electromagnetic field
observable. However, the regular Weyl operators, as discussed
earlier, are functions of the free outgoing field only, and the
Coulomb field is lost. If the vacuum is of the form (ii) of the
theorem, one can separately {\em define} it by the first term in
(\ref{Gauq}), but this is done ``by hand'', and the information on
the unique way in which the Coulomb field and the ``out'' field add
to form the total field is lost.\\
(ii) The superselection sectors with respect to regular observables
$\A_0\df\F_0\cap\A$ are labeled by the spacelike asymptotic of the
{\em free} field (and, of course, by total charge, if it exists), so
that even the standard wisdom does not apply here. The spacelike
asymptotic of electromagnetic field according to Buchholz \c{buch} 
yields in the subspace $[\pi(W(V))\pi(\F_0)\W]$ the field
(\ref{spas}), which is a free field for free field asymptotic $V$. 
Moreover, $\A_0$ should not be interpreted as the algebra of local
observables: creation or annihilation of a charged particle together
with its Coulomb field is a nonlocal operation, so
$B(g)^*B(f)\in\A_0$ is a nonlocal observable. 

Consider the particular, Poincar\'e-invariant vacuum state as given
by Theorem \ref{vac1}(ii) with $f\equiv 1$. In this representation
there is no infraparticle problem: all one-particle states
$\pi(B(g_+)^*)\W$, $\pi(B(g_-))\W$ have energy-momentum on the mass
hyperboloid. We want to construct
representations which remain as close in their structure to this
vacuum representation, but which are regular at the same time. The
obvious idea how to do it, is to try to integrate the superselection
sectors of this vacuum into a direct integral Hilbert space. As the
representation is of the form determined by Theorem \ref{tcov}, it is
sufficient to confine attention to the electromagnetic part of the
representation, $\pi_r$. However,
for measure-theoretical reasons one has to extend the scope of
sectors which are to be integrated. It is now that the extension of
the symplectic space introduced in Sec.\ref{auw} will be needed. This
extension allows us to consider the Weyl algebra $\Fa$ generated
uniquely (due to nondegeneracy of the symplectic form \c{mstv}) by
elements $\hW([u])$, $[u]\in\La$, 
$$
\hW([u_1])\hW([u_2]) = e^{\txt -\frac{i}{2}\{[u_1], [u_2]\} } 
\hW([u_1+u_2])\, .
$$
The elements $\hW([u])$ with $[u]$ from  the
subspace $\La_0$ generate a $C^*$-subalgebra, which we denote
$\Fa_0$. The Poincar\'e transformations act on the algebra $\Fa$ by 
\begin{eqnarray*}
&&\a_{x,A}\hW([u]) =\hW([T_{x,A}u])\, \\
&&\lp T_{x,A}u\rp_a(s,l)=\Lambda(A)_a{}^b\, u_b(s-x\s
l,\Lambda^{-1}l)\, .
\end{eqnarray*}
The Poincar\'e invariant vacuum state is easily obtained:
$$
\hat{\w}(\hW([u])) =\left\{
\begin{array}{ll}
\dsp 0\, , &\mbox{if}~~p([u])\neq 0\, ,\\
\dsp e^{\txt -\frac{1}{2} F([u], [u])}\, , &\mbox{if}~~p([u])=0\, .
\end{array}
\right.
$$
The representation space $\hH$ of the representation $\hpi$
canonically obtained from $\hat{\w}$ is the uncountable direct sum 
$\dsp \hH=\bigoplus_{[f]\in\HI} \Hf$, where $\Hf$ is the
subspace spanned by the vectors $\hpi(\hW([u]))\hO$ with
$p([u])=[f]$. The restriction of the representation $\hpi$ to the
subalgebra $\Fa_0$ and to the subspace $\Hf$ is a
coherent state representation \c{roe70}, 
$\hpi_{[f]}(\Fa_0)\df \hpi(\Fa_0)\big|_{\hH_{[f]}}$. 
Two representations $\hpi_{[f_i]}$, $i=1,2$, are disjoint for
$[f_1]\neq[f_2]$ \c{roe70}. In particular,
$\hpi_{[0]}(\Fa_0)$ is the (positive energy) Fock
representation. Each subspace $\Hf$ is invariant under the
action of translations $\hU(x)$, thus
$\dsp \hU(x)=\bigoplus_{[f]\in\HI} \hU_{[f]}(x)$, each component
representation being a positive energy representation, 
$\mbox{Spec}\hU_{[f]}(x)\subset\ov{V_+}$. 

Suppose we are given a cylidrical $\sigma$-additive measure $\mu$ on
the Hilbert space $\HI$ (\ref{IR}) \c{gel}. Let $\p([f])$ vary over
the set of all measurable functions $\HI\rightarrow\hH_{[0]}$ (i.e.,
the set of functions for which $(\ph, \p([f]))_{\hH_{[0]}}$ is
measurable for all $\ph\in\hH_{[0]}$ \c{dix}). 
Let further $h(s,l)$ vary over
the set of smooth homogeneous functions of degree $-1$, satisfying 
for some $\e>0$ and all $k=0,1,\ldots$ the bounds 
\be
| L_{b_1c_1}\ldots L_{b_kc_k} h(s,l)|
\leq\frac{\con(k)}{(|s|+1)^{1+\e}}
\label{boh}
\ee
(for any gauge $t\s l=1$) and such that $\ti{h}(0,l)=1$. 
Then $[fh]\in \La$ and $p([fh])=[f]$ for any $[f]\in\HI$.
Consider the set $\Gamma$ of all functions of the form 
\be
\HI\ni [f]\rightarrow\Psi([f])=\hpi(\hW([fh]))\p([f])\in\Hf\, .
\label{ga1}
\ee
\begin{lem}
For any fixed $h$ every function in $\Gamma$ has a unique
representation (\ref{ga1}). 
\label{ga1'}
\end{lem}
Proof. If $h'$ is another function satisfying the same conditions as
$h$ then 
$$
\hpi(\hW([fh']))\p([f]) 
=\hpi(\hW([fh]))e^{\txt \frac{i}{2}\{ [fh], [fh']\} }
\hpi_{[0]}(\hW([f(h'-h)])) \p([f])\, .
$$
It has to be shown that if $\p([f])$ is measurable, then 
$\p'([f])=\\
e^{\txt \frac{i}{2}\{ [fh], [fg]\}}
\hpi_{[0]}(\hW([fg]))\p([f])$, where $g\equiv h'-h$, is also
measurable. It is sufficient to show 
that $(e_i, \p'([f]))$ are measurable for a
basis $\{ e_i\}$ of $\hH_{[0]}$. If one chooses for $\{ e_i\}$ the
finite ``particle'' number basis, then these products have the
following form: $\dsp e^{\txt \frac{i}{2}\{ [fh], [fg]\}
-\frac{1}{2}F([fg], [fg])} \sum_{k=1}^\infty (e_k, \p([f]))
C_k([f])$, where $C_k([f])$ are linear
combinations of products of expressions $F([fg], [\chi])$ and their
complex conjugations, where $[\chi]\in \La_0$ are profiles of the
photons in the basis vectors. Now, the expressions $F([fg], [\chi])$,
$\{ [fh], [fg]\}$, and $F([fg], [fg])$ are easily seen to be
absolutely bounded as functions of $[f]$, the first by
$\con\|[f]\|_{\HI}$, the other two by $\con\|[f]\|^2_{\HI}$. Hence, 
the first may be written as $([f], [k])$, $[k]\in\HI$, and
the other two as $([f], B[f])_{\HI}$, where $B$ is a bounded operator
on $\HI$. Both expressions are measurable functions of $[f]$, which
ends the proof. \ep

The set $\Gamma$ thus has the natural structure of a vector space,
and the pair $\lp\{\Hf\}_{[f]\in\HI}, \Gamma\rp$ is easily seen to form a
$\mu$-measurable family of Hilbert spaces \c{dix}. This family
determines the direct integral Hilbert space \\
$\dsp\H=\int_{\oplus}\Hf\, \m([f])$, with its elements denoted by 
$\dsp\Psi=\int_{\oplus}\Psi([f])\, \m([f])$. 
It is necessary for our purposes to assume that $\mu$ is
quasi-invariant with respect to translations by smooth elements of
$\HI$. This set, more exactly the set of elements in $\HI$ having
smooth representants, will be denoted by $\CIR$. Hence, we demand
that  
\be
\begin{array}{@{}l}
\dsp \mu_{[k]}(B)\df \mu(B-[k])=0~~\iff~~\mu(B)=0\\
\dsp \forall~\mbox{measurable}~B\, ,~[k]\in\CIR\, .
\end{array}
\label{qi1}
\ee
(This was the reason for the need to extend $L^f$ to $\La$, and
consequently $\CIR$ to $\HI$: there are no measures on function
spaces quasi-invariant with respect to all translations \c{gel}.) 
Then, by Radon--Nikodym theorem, 
$\m_{[k]}([f])=\lp\m_{[k]}/\m\rp([f])\,\m([f])$, where 
$\lp\m_{[k]}/\m\rp([f])$ and $\lb\lp\m_{[k]}/\m\rp([f])\rb^{-1}$ 
are integrable, non-negative functions. 

Formula (\ref{ga1}) may be used to define, for each fixed $h$, an
isomorphism of the space $\H$ with the tensor product space 
$L^2(\HI, \mu)\dip \hH_{[0]}$. Namely, we put by definition
\be
\begin{array}{@{}l}
\dsp \U_h: L^2(\HI, \mu)\dip \hH_{[0]}\rightarrow \H\, ,\\
\dsp \chi\dip\p\rightarrow\Psi=\int_{\oplus}\Psi([f])\, \m([f])\, ,\\
\dsp \Psi([f])=\hpi(\hW([fh]))\chi([f])\p\, .
\end{array}
\label{sup4}
\ee
This is a noncanonical isomorphism, as it depends on the choice of
the function $h$, which has no intrinsic meaning. The use of the
isomorphism will be restricted to technical purposes only. 

We now define in $\H$ a new representation of CCR. Note that if $V$
is a test function of $W(V)$, then by (\ref{dec}) $\tV(\w, l)$ is a
smooth function of $l$ for each $\w$. Therefore $p([\V])$ is a smooth
element of $\HI$ (has a smooth representant $\tV(0, l)$). 
\begin{theor}
The linear operators on $\H$ introduced by
\be
\begin{array}{@{}l}
\dsp [\pi_r(W(V))\Psi]([f])\df\\
\dsp \lp\frac{\m_{p([\V])}}{\m}([f])\rp^{\frac{1}{2}}\hpi(\hW([\V])) 
\Psi([f]-p([\V]))\, ,
\end{array}
\label{sup1}
\ee
define a representation $\pi_r$ of {\rm CCR} satisfying Borchers' 
criterion. The unitary representation of translations defined by 
\be
[U_r(x)\Psi]([f]) \df\hU_{[f]}(x)\Psi([f])
\label{sup2}
\ee
implements translations in the representation $\pi_r$ and satisfies 
${\rm Spec}U(x) =\ov{V_+}$. 
\label{sup0}
\end{theor}
Proof. A straightforward calculation shows that $\pi_r$ and $U_r$
satisfy the algebraic conditions of representations and that $U_r(x)$
implement translations in the representation $\pi_r$. 
From the strong continuity and spectral properties of each
of $\hU_{[f]}(x)$ it follows that also $U_r(x)$ is strongly continuous
and $\mbox{Spec}U_r(x)\subset\ov{V_+}$. The proof will be now
completed by showing that ${\rm Spec}(x)\hU_{[0]}\subset {\rm
Spec}U_r(x)$ ($\hU_{[0]}$ is the standard transformation of the Fock
representation, hence ${\rm Spec}(x)\hU_{[0]}= \ov{V_+}$). 
Take $\Psi =\U_h \lp\chi\dip\p\rp$, with $\|\chi\|_{L^2(\HI,\mu)}=1$,
$\|\p\|_{\hH_{[0]}}=1$. Then 
\begin{eqnarray*}
&&\dsp (\Psi, U_r(x)\Psi)\\
&&\dsp =\int|\chi([f])|^2e^{\txt \frac{i}{2} 
\{[fh], [fT_xh]\} } \lp\p, \hpi_{[0]}(\hW([f(T_xh-h)])) 
\hU_{[0]}(x)\p\rp\, \m([f])\, .
\end{eqnarray*}
Let $\Psi_k=\U_{h_k}\lp\chi\dip\p\rp$ ($k=1,2\ldots$), with 
$h_k(s,l)= k^{-1} h(k^{-1}s,l)$. Then\\
$\dsp\lim_{k\to\infty} F([fh_k], [f(T_xh_k-h_k)])=0$,
$\dsp\lim_{k\to\infty} F([f(T_xh_k-h_k)], [f(T_xh_k-h_k)])=0$\\ and 
$\dsp {\rm w-}\!\lim_{k\to\infty} \hpi_{[0]}\lp\hW([f(T_xh_k-h_k)])\rp 
=1$. By Lebesgue's theorem \\
$\dsp\lim_{k\to\infty}
(\Psi_k, U_r(x)\Psi_k)= (\p, \hU_{[0]}(x)\p)$. 
Hence ${\rm Spec}\hU_{[0]}(x)\subset {\rm Spec}U_r(x)$.   \ep

Further properties of the representation $\pi_r$ are discussed 
after finding its unitarily equivalent form $\U_h^{-1}\pi_r\U_h$ by
(\ref{sup4}). Let $V_a(s,l)$ be any test function of $W(V)$. The
integral 
$$
-\frac{1}{4\pi}\int h(s,l)\e(s-\tau)\V_a(\tau,l)\, ds\, d\tau= 
-i\, \mbox{P}\int\ov{\ti{h}(\w, l)}\tV_a(\w, l)\,
\frac{d\w}{\w}
$$ 
is a real smooth vector function orthogonal to $l$, and a
representant of an element of $\H_0$ (defined before (\ref{gen3})).
The orthogonal projection of this element to the subspace $\HI$ of
$\H_0$ will be denoted by $r_h([\V])$. We show in the Appendix that
the orthogonal projection to $\HI$ of a smooth element of $\H_0$ is
smooth, hence $r_h([\V])\in\CIR$. Let ccr be the Weyl 
algebra over the vector space $\CIR\dis\CIR$ with the symplectic form
$\{[g_1]\dis[k_1], [g_2]\dis[k_2]\}_{IR}\df ([g_1], [k_2])_{\HI} -
([k_1], [g_2])_{\HI}$. This algebra is generated by elements 
$w([g]\dis [k])$ satisfying 
\be
\begin{array}{@{}l}
w([g_1]\dis [k_1]) w([g_2]\dis [k_2])\\
~~=e^{\txt -\frac{i}{2} \{[g_1]\dis[k_1], [g_2]\dis[k_2]\}_{IR} } 
w([g_1+g_2]\dis [k_1+k_2])\, ,\\
w([g]\dis [k])^*=w(-([g]\dis [k]))\, .
\end{array}
\label{ccr}
\ee
(This algebra is unambiguously defined as the symplectic form is
nondegenerate \c{mstv}.) 
\begin{theor} \hspace*{\fill}\\
(i) The following definition determines a representation $\pi_\mu$ of
the algebra {\rm ccr} on the Hilbert space $L^2(\HI, \mu)$
\be
\begin{array}{@{}l}
\dsp \lp\pi_\mu(w([g]\dis [k]))\chi\rp ([f])\df \\
\dsp \lp\frac{\m_{[g]}}{\m}([f])\rp^{\frac{1}{2}} 
e^{\txt i([f]-\frac{1}{2}[g], [k])_{\HI}} \chi([f-g])\, .
\end {array}
\label{sup6}
\ee
The vector $\W_\mu([f])=1$ is cyclic in $L^2(\HI, \mu)$ for the
representation $\pi_\mu$ restricted to the subalgebra generated by
elements $w([0]\dis [k])$. \\
(ii) The representation $\pi_r$ goes over under the unitary
transformation $\U_h$ of representation space to
\be
\U^{-1}_h\pi_r(W(V))\U_h = \pi_\mu(w(p([\V])\dis r_h([\V])))
\dip\hpi_{[0]}(\hW([\V-\tV(0,.)h]))\, .
\label{sup7}
\ee
(iii) $(\U_h^{-1}\pi_r({\rm CCR})\U_h)''= \pi_\mu({\rm ccr})''\dip 
{\cal L}(\hH_{[0]})$ (the von Neumann tensor product), where 
${\cal L}(\hH_{[0]})$ is the algebra of bounded operators on 
$\hH_{[0]}$. Therefore, $\pi_r$ is irreducible iff 
$\pi_\mu$ is irreducible. \\
(iv) $\pi_r$ is regular iff $\pi_\mu$ is regular.
\label{sup5}
\end{theor}
Proof. \hspace*{\fill}\\
(i) The algebraic properties needed for $\pi_\mu$ to be a
representation of ccr are checked by direct calculation. For any
$\chi\in L^2(\HI, \mu)$ there is 
$$
\lp\chi, \pi_\mu(w([0]\dis [k]))\W_\mu\rp= 
\int\ov{\chi([f])}e^{\txt i([f], [k])}\, \m([f])\, ,
$$
If this integral vanishes for all smooth $[k]$, then by continuity in
$[k]$ it vanishes for all $[k]\in\HI$ and then $\chi=0$, which shows
that $\W_\mu$ is cyclic.\\
(ii) A direct calculation gives 
\begin{eqnarray*}
&&\lb\pi_r(W(V))\U_h\lp\chi\dip\p\rp\rb([f]) =\hpi(\hW([fh]))
\lp\frac{\m_{p([\V])}}{\m}([f])\rp^{\frac{1}{2}}\times\\
&&e^{\txt i\{[(f-\frac{1}{2}\tV(0,.))h], [\V]\} } 
\chi([f]-p([\V]))\hpi_{[0]}(\hW([\V-\tV(0,.)h]))\p\, .
\end {eqnarray*}
Denote $f'\equiv f-\frac{1}{2}\tV(0,.)$. The symplectic form in the
exponent of the last formula is evaluated by
\begin{eqnarray*}
&&\{[f'h], [\V]\}=-\int f'_a(l)\lp -i\, \mbox{P}
\int\ov{\ti{h}(\w,l)}\tV^a(\w,l)\, \frac{d\w}{\w}\rp\, \dl\\
&&=([f'], r_h([\V]))_{\HI}= ([f] - \frac{1}{2}p([\V]),
r_h([\V]))_{\HI}\, .
\end{eqnarray*}
Transforming back with $\U_h^{-1}$ one obtains (\ref{sup7}). \\
(iii) Let $(V')\in L_0$. Further, let $f_a(l)$ be a smooth function
representing an element $[f]\in\HI$ and put $\V_a(s,l)=
f_a(l)h(s,l)+\V'_a(s,l)$. Then $(V)\in L$, $p([\V])=[f]$, and 
\be
\U_h^{-1}\pi_r(W(V))\U_h = \pi_\mu(w([f]\dis r_h([\V'])))
\dip\hpi_{[0]}(\hW([\V']))\, .
\label{clos1}
\ee
The functions $\ti{\V'}$ are dense in the Hilbert space of test
functions of the Fock representation, which is irreducible.
Therefore, to prove the equality in (iii) it is sufficient to show
that all operators  $\pi_\mu(w([f]\dis [g]))\dip \I_{[0]}$ are in\\
$(\U_h^{-1}\pi_r({\rm CCR})\U_h)''$. The second statement of (iii)
then follows from  
$(\U_h^{-1}\pi_r({\rm CCR})\U_h)'=\pi_\mu(\mbox{ccr})'\dip 
\C\I_{[0]}$ (see \c{dix}). \\
To fill the missing step consider for $\b\in(0, \e)$ a one-parameter
family of functions 
$\ti{N_\b}(\w,l)=\ti{\k}_\b(\w t\s l)\ti{h}(\w,l)$, 
where $\ti{\k}_\b(\w)=ie^{-|\w|}|\w|^\b\,{\rm sgn}(\w)$. Then
$N_\b(s,l)$ are real homogeneous functions of degree $-1$ given by\\ 
$\dsp N_\b(s,l)=\frac{1}{2\pi t\s l}
\int\k_\b\lp\frac{s-\tau}{t\s l}\rp
h(\tau,l)\, d\tau$, with \\
$\dsp\k_\b(s)=2\Gamma(1+\b) 
(s^2+1)^{-(1+\b)/2} \sin{((1+\b)\mbox{arctg}s)}$. With the use of
bounds (\ref{boh}) one shows that $N_\b(s,l)$ are smooth and also
satisfy bounds of the form (\ref{boh}), with $\e$ replaced by
$\b$. Consider the smooth homogeneous function of degree $0$ given by
the integral 
$$
c_\b(l)= -\frac{1}{4\pi}\int h(\tau, l)\e(\tau-s)N_\b(s, l)\, 
d\tau\, ds 
= 2\int_0^{\infty}\left|\ti{h}\lp\frac{\w'}{t\s l},l\rp 
\right|^2 e^{-\w'}\w'^{\b-1}\, d\w'\, ,
$$
the last equality by (\ref{P1}). From the bound of the form
(\ref{norm4}) satisfied by $\ti{h}$ and the condition $\ti{h}(0,
l)=1$ we know that there is a positive $u$, such that 
$\left|\ti{h}(\w'/t\s l,l)\right|^2 e^{-\w'}>
\frac{1}{2}$ for $\w'\in\langle 0, u\rangle$. On the other hand, the
function $\left|\ti{h}(\w'/t\s l,l) \right|^2$ is
bounded by a constant from above. Hence,
$(u^\b/\b) < c_\b(l) < \con \Gamma(\b)$. 
Therefore, the new auxiliary function defined by 
$n_\b(s,l)=N_\b(s,l)/c_\b(l)$ is smooth, has all the
properties listed above for $N_\b(s,l)$, and in addition satisfies 
\be
-\frac{1}{4\pi}\int h(\tau, l)\e(\tau-s)n_\b(s, l)\, 
d\tau\, ds=1\, .
\label{en}
\ee
Now, choose a smooth element $[g]$ in $\HI$ and put
$\V'_{\b a}(s,l)=g_a(l)n_\b(s,l)$. These $V'_\b$ satisfy the
conditions for a test function $V'$ in (\ref{clos1}) and, by
(\ref{en}), yield $r_h([\V'_\b])= [g]$. For 
$\V_{\b a}(s,l)=f_a(l)h(s,l)+\V'_{\b a}(s,l)$ Eq.(\ref{clos1}) reads 
\be
\U_h^{-1}\pi_r(W(V_\b))\U_h = \pi_\mu(w([f]\dis [g]))
\dip\hpi_{[0]}(\hW([\V'_\b]))\, .
\label{clos4}
\ee
By a straightforward calculation one obtains 
\begin{eqnarray*}
&&F([\V'_\b], [\V'_\b]) = \int \dl\, \frac{-g^2(l)}{c_\b^2(l)} 
\int_0^\infty \left|\ti{h}\lp\frac{\w'}{t\s l},l\rp 
\right|^2 e^{-2\w'}\w'^{2\b-1}\, d\w'\\ 
&&< \int (-g^2(l))\frac{c_{2\b}(l)}{2c_\b^2(l)}\,\dl <
\con\frac{\b^2\Gamma(2\b)}{u^{2\b}}\int \lp -g^2(l)\rp\, \dl 
=\con\frac{\Gamma(1+2\b)}{u^{2\b}} \b\, .
\end{eqnarray*}
Hence $\dsp\lim_{\b\to 0} F([\V'_\b], [\V'_\b]) =0$, and then 
$\dsp {\rm w-}\!\lim_{\b\to 0} \hpi_{[0]}(\hW([\V'_\b])) =\I$. 
Then, by (\ref{clos4}), 
$\dsp {\rm w-}\!\lim_{\b\to 0} \U_h^{-1}\pi_r(W(V_\b))\U_h = 
 \pi_\mu(w([f]\dis [g]))\dip \I$, which ends the proof of (iii).\\
(iv) This is obvious by regularity of the Fock representation and by 
$$
\U^{-1}_h\pi_r\lp W((\lambda V)_\Phi)\rp\U_h = 
\pi_\mu(w(\lambda (p([\V])\dis r_h([\V]))))
\dip\hpi_{[0]}(\hW(\lambda [\V-\tV(0,.)h]))\, .
$$  \ep

Remarks. (i) Theorems \ref{tcov}, \ref{sup0}, and \ref{sup5} together
characterize 
a class of regular, irreducible representations of the algebra $\F$ 
satisfying Borchers' criterion. When restricted to ${\rm
CCR}\cap\F_0$ the representations decompose into direct integral of
coherent state representations and in this respect resemble the
scattering representations considered in Ref.\c{fms}. However, here
the infrared clouds are independent of the charged particles (they
are there even if there are no such particles present). In
particular, the arguments of this reference for the Lorentz symmetry
breaking do not apply here. \\
(ii) The vacuum vector is replaced here by
``infravacua'', and states with finite charged particle number are
obtained by the action of creation operators on any such state. 
(The ``infravacua'' are not of the KPR type \c{kpr}, which does not
lead to coherent states.) There are no vectors with the
energy-momentum on mass hyperboloid, but the arguments of
Ref.\c{buch} do not apply here either: the asymptotic of
electromagnetic field according to Buchholz catches only the free
field part, and does not characterize states by classical
distribution of electric flux. \\
(iii) The operators $\pi(W(V))$ with $l\wedge V(-\infty,l)\neq 0$
representing the exponentials of total 
electromagnetic field do not commute with $\pi(B(g))$, which reflects
the fact that creation or annihilation of a charged particle together
with its Coulomb field is a nonlocal operation. They do not commute
with nonlocal observables $\pi(B(g)^*B(f))$ either. 

A particular representation in the class thus characterized is
given whenever a measure $\mu$ is chosen, such that the
condition (\ref{qi1}) is satisfied and the representation $\pi_\mu$
is regular and irreducible. Explicit characterization of such
measures may be given in the subclass of Gaussian measures. For any
positive, trace-class operator $B$ in the Hilbert space $\HI$ the
characteristic function 
$$
\int e^{\txt i([f], [g])}\,\m_B([f])= e^{\txt -\frac{1}{2} 
([g], B[g])}
$$
defines a cylindrical, $\sigma$-additive measure, a Gaussian measure
with covariance $B$ \c{df}. The following proposition is obtained by
the application of general standard results. 
\begin{pr}  \hspace*{\fill}\\
(i) A Gaussian measure with covariance $B$ satisfies the
quasi-invariance condition (\ref{qi1}) iff 
\be 
\CIR\subset B^{\frac{1}{2}}\HI
\label{gau2}
\ee 
(hence ${\rm Ker}B^{\frac{1}{2}}=
\{ 0\}$, as $\ov{\CIR}^{\HI}=\HI$). Then the representation $\pi_\mu$
is regular and it is unitarily equivalent to the GNS representation
of the quasi-free state 
\be
\w_\mu(w([g]\dis [k]))= e^{\txt -\frac{1}{4}s([g]\dis [k], [g]\dis
[k])} \, ,
\label{gau3}
\ee
where $s$ is a bilinear, positive definite form on $\CIR\dis\CIR$
satisfying the defining condition of a quasi-free state 
\be
|\{ [g_1]\dis [k_1], [g_2]\dis [k_2]\}_{IR}|^2\leq 
s([g_1]\dis [k_1], [g_1]\dis [k_1]) 
s([g_2]\dis [k_2], [g_2]\dis [k_2]) 
\label{gau4}
\ee
and given explicitly by 
\be
s([g_1]\dis [k_1], [g_2]\dis [k_2]) = 
\frac{1}{2}\lp B^{-\frac{1}{2}}[g_1], B^{-\frac{1}{2}}[g_2]\rp + 
2\lp B^{\frac{1}{2}}[k_1], B^{\frac{1}{2}}[k_2]\rp\, .
\label{gau5}
\ee
(ii) If (\ref{gau2}) is satisfied, then $\pi_\mu$ is irreducible iff 
$\ov{B^{-\frac{1}{2}}\CIR}^{\HI}=\HI$. The state (\ref{gau3}) then
yields a Fock representation. Otherwise $\pi_\mu$ is non-factor. 
\label{gau6}
\end{pr}
Proof. 
(i) A Gaussian measure on a Hilbert space $\HI$ and with covariance
$B$ is equivalent to its translation by an element $[g]\in\HI$ if,
and only if, $[g]\in B^{\frac{1}{2}}\HI$ (see \c{df}), which proves
the first statement of (i). If this is the case, then 
$$
\frac{\m_{[g]}}{\m}([f])= 
e^{\txt -\frac{1}{2} \| B^{-\frac{1}{2}}[g]\|^2 + 
([f], B^{-1}[g])}\, ,
$$
where $([f], B^{-1}[g])$ is to be understood in the sense of a
measurable linear functional on $\HI$ \c{df}. As the vector
$\W_{\mu}$ is cyclic for $\pi_\mu$, this representation is unitarily
equivalent to the GNS representation obtained from the state 
$\w_\mu(A)\df (\W_\mu, \pi_\mu(A)\W_\mu)$. The relations
(\ref{gau3})--(\ref{gau5}) are now verified by calculation. By the
results of Ref.\c{mv} $\w_\mu$ is a quasi-free state, which always
yields a regular representation. \\
(ii) By the results of \c{mv} the state $\w_\mu$ is primary iff 
the extension $\{.,.\}_{IR}^s$ of the symplectic form 
$\{.,.\}_{IR}$ to the completion of $\CIR\dis\CIR$ with respect to
$s$ is nondegenerate. This completed space is a real Hilbert space
$\H_s$. In our case $\H_s=\H_-\dis\H_+$, where $\H_-$ (resp. $\H_+$)
is the completion of $\CIR$ with respect to the norm 
$\|[f]\|_-\df \| B^{-\frac{1}{2}}[f]\|$ (resp. 
$\|[f]\|_+\df \| B^{\frac{1}{2}}[f]\|$), and $s$ is extended to 
$\dsp (x_1\dis y_1, x_2\dis y_2)=\frac{1}{2}(x_1, x_2)_- +2(y_1,
y_2)_+$. The linear operator $U_-$ (resp. $U_+$) on $\CIR$ defined by
$U_-[f]\df B^{-\frac{1}{2}}[f]$ 
(resp. by $U_+[f]\df B^{\frac{1}{2}}[f]$) maps $\CIR$ as a subspace
of $\H_-$ (resp. of $\H_+$) isometrically into $\HI$. By continuity 
$U_\mp$ extend to isometric operators $U_\mp:\H_\mp\rightarrow\HI$. 
By restricting arguments in the following equation to smooth elements
one sees that the extension of $\{.,.\}_{IR}$ to $\{.,.\}_{IR}^s$ is
given by $\{ x_1\dis y_1, x_2\dis y_2\}_{IR}^s = (U_-x_1,U_+y_2)_{IR}
- (U_-x_2, U_+y_1)_{IR}$. This form is nondegenerate iff 
$U_-\H_-=U_+\H_+$. However,
$U_+\H_+=\ov{B^{\frac{1}{2}}\CIR}^{\HI}=\HI$ (as ${\rm
Ker}B^{\frac{1}{2}}=\{ 0\}$). Therefore, $\w_\mu$ is primary iff 
$U_-\H_-\equiv\ov{B^{-\frac{1}{2}}\CIR}^{\HI}=\HI$. If this is the
case, then it is checked by direct calculation that 
$(x_1\dis y_1, x_2\dis y_2)_s=\{ x_1\dis y_1, A(x_2\dis
y_2)\}^s_{IR}$,  
where $A(x\dis y)\df \lp -2U_-^{-1}U_+ y\rp\dis 
\lp\frac{1}{2} U_+^{-1}U_- x\rp$. The operator $A$ satisfies the
equation $A^2=-1$, which is a necessary and sufficient condition for
$\w_\mu$ to be pure \c{mv}. \ep

Concrete examples of trace-class operators $B$ satisfying the
conditions of Proposition \ref{gau6} (i) and (ii) are most easily
constructed in the unitarily equivalent version of the space $\HI$,
the Hilbert space $\H_{\D^2}$ discussed in the Appendix. Let 
$\ti{B}_\a=\lb\lp(t\s l)^2\D^2\rp^{-1}\rb^{1+\a}$, $\a>0$, where 
$\lp(t\s l)^2\D^2\rp^{-1}$ is the positive operator on $\H_{\D^2}$
defined in Appendix (Eq.(\ref{ras}) and the following discussion). It
follows from the spectral properties of this operator that each of
the operators $\ti{B}_\a$ may serve as an example of the transformed
covariance operator. 

\section*{Acknowledgements}
\label{ack}

I am grateful to Professor Andrzej Staruszkiewicz for discussions and
support, to Professor Detlev Buchholz for helpful correspondence, and
to Walter Kunhardt for discussions and critical reading of an early
version of the manuscript. 

\setcounter{equation}{0}
\setcounter{pr}{0}
\setcounter{defin}{0}
\setcounter{section}{0}
\renewcommand{\thesection}{Appendix}
\renewcommand{\theequation}{\Alph{section}.\arabic{equation}}
\renewcommand{\thepr}{\Alph{section}.\arabic{pr}}

\section{\hspace*{-.7cm}: Homogeneous functions on the lightcone}
\label{app}

In this Appendix we briefly discuss some structures and
operations on spaces of homogeneous functions on the future
lightcone. Let, first, $\phi(l)$, $\phi_1(l)$ and $\phi_2(l)$ be
smooth ($\Ci$ in the sense of differentiation by
$L_{ab}=l_a\D_b-l_b\D_a$) functions, homogeneous of degree $0$. Take,
for the sake of differentiation, extensions of these functions which
remain homogeneous in some neighbourhood of the cone. Straightforward
calculation then gives on the cone 
\be
L_a{}^b L_{cb}\phi= {}^*L_c{}^b{}^*L_{ab}\phi= -L_{ac}\phi
+l_al_c\,\D^2\phi\, ,
\label{a1}
\ee
\be
{}^*L_a{}^b L_{cb}\phi= -L_c{}^b{}^*L_{ab}\phi= -{}^*L_{ac}\phi\, ,
\label{a2}
\ee
\be
L_a{}^b\phi_1 L_{cb}\phi_2= {}^*L_c{}^b\phi_1{}^*L_{ab}\phi_2 
= l_al_c\, \D\phi_1\s \D\phi_2\, ,
\label{a3}
\ee
where ${}^*L_{ab}=\frac{1}{2}e_{abcd}L^{cd}$ and $\D^2=\D_a\D^a$. As
the action of $L_{ab}$ is extension-independent (this is the tangent
derivative), these formulae give extension-independent meaning of
$\D^2\phi$ and $\D\phi_1\s \D\phi_2$, which in this form were
calculated for a homogeneous (but otherwise arbitrary) extensions. 
Contracting (\ref{a1}) and (\ref{a3}) with $t^at^c$, where $t$ is any
unit future-pointing vector, one obtains 
\be
\D^2\phi = (t\s l)^{-2} {}^*L_0{}^b {}^*L_{0b}\phi\, ,
\label{a4}
\ee
\be
\D\phi_1\s \D\phi_2 = 
(t\s l)^{-2} {}^*L_0{}^b\phi_1 {}^*L_{0b}\phi_2\, ,
\label{a5}
\ee
where ${}^*L_{0b}=t^a{}^*L_{ab}$. If $\p$ is any differentiable
function, homogeneous of degree $-2$, then 
\be
\int L_{ab}\p(l)\, \dl =0\, .
\label{a6}
\ee
Using this identity one obtains from (\ref{a4}) and (\ref{a5}) by
integration by parts (and taking into account ${}^*L_{0b}t\s l=0$) 
\be
\int \phi_1\D^2\phi_2\, \dl = -\int \D\phi_1\s \D\phi_2\, \dl\, .
\label{a7}
\ee
Therefore, $\int\phi\,\D^2\phi\, \dl\geq 0$ and 
$\int\phi\,\D^2\phi\, \dl= 0$ iff $\phi=\con$. Thus $\D^2\phi=0$
iff $\phi=\con$. This positivity of $\D^2$ is also seen from the
identity (\ref{a4}), which says that $\D^2$, when applied to a
homogeneous function of degree $0$, is the ``orbital angular momentum
squared'' in each Minkowski frame. 

The action of $\D^2$ may be explicitly reversed. For each smooth,
homogeneous of degree $0$ function $\phi$ the function $\p=\D^2\phi$
is smooth, homogeneous of degree $-2$, and satisfies 
$\int\p\,\dl=0$. Conversely, if $\p$ is any function with these
properties, then the formula 
\be
\phi_t(l)= -\frac{1}{4\pi}\int\mbox{ln}\frac{l\s l'}{t\s l'}\p(l')\,
\dl' 
\label{a8}
\ee
gives the unique such smooth function that $\D^2\phi_t=\p$, and with
the additive constant chosen such that 
\be
\int \frac{\phi_t(l)}{(t\s
l)^2}\, \dl=0\, . 
\label{a8'}
\ee
Smoothness of $\phi_t(l)$ is proved by showing that
for $\e\searrow 0$ the functions 
$$
\phi_{t\e}(l)=-\frac{1}{4\pi}\int\mbox{ln}\frac{l\s l'+\e t\s lt\s
l'}{t\s l'}\p(l')\, \dl'
$$
converge uniformly to $\phi_t(l)$, and $L_{ab}\phi_{t\e}(l)$
converge uniformly to 
$$
-\frac{1}{4\pi}\int\mbox{ln}\frac{l\s l'}{t\s l'}L'_{ab}\p(l')\, \dl'
+ \frac{1}{4\pi}\int\frac{l'_at_b-l'_bt_a}{t\s l'}\p(l')\,\dl'\, .
$$
Then it remains to show that 
\begin{eqnarray*}
&&-\frac{1}{4\pi(t\s l)^2}\int \mbox{ln}\frac{l\s l'+\e t\s lt\s
l'}{t\s l'} {}^*L'_0{}^b{}^*L'_{0b} \p(l')\, \dl'\\
&&=\frac{2\e+\e^2}{4\pi}\int
\lp\frac{t\s l'}{l\s l'+\e t\s lt\s l'}\rp^2\p(l')\,\dl'
\end{eqnarray*}
converges to $\p(l)$ pointwise, which is an easy exercise. 

Consider now the linear space $[\Ci]$ of equivalence classes
$[\phi]=\{\phi'|\, \phi'=\phi+\con\}$ of smooth, homogeneous of
degree $0$, functions $\phi$. On this space the operators 
$[\phi]\rightarrow [(t\s l)^2\D^2\phi]$ and 
$[\phi]\rightarrow [((t\s l)^2\D^2)^{-1}\phi_t]$, where $\phi_t$ is
the representant satisfying (\ref{a8'}), and 
\be
\lp ((t\s l)^2\D^2)^{-1}\phi_t\rp\! (l)=-\frac{1}{4\pi}\int\mbox{ln}
\frac{l\s l'}{t\s l'}\, \frac{\phi_t(l')}{(t\s l')^2}\, \dl'\, ,
\label{ras}
\ee
are bijective, mutually inverse maps. 
Moreover, it is easily seen that the operator $((t\s l)^2\D^2)^{-1}$
is bounded with respect to the norm of the scalar product on $[\Ci]$ 
defined by 
\be
([\phi_1], [\phi_2])_t = \int\frac{\phi_{1t}\phi_{2t}(l)}{(t\s
l)^2}\, \dl\, ,
\label{a9}
\ee
that is 
\be 
\left\|\lb((t\s l)^2\D^2)^{-1}\phi_t\rb\right\|_t\leq 
c\|[\phi_t]\|_t\, .
\label{a10}
\ee
The expression (\ref{a7}) introduces another scalar product on
$[\Ci]$ 
\be
([\phi_1], [\phi_2])_{\D^2} =\int\phi_1\D^2\phi_2(l)\, \dl\, .
\label{a11}
\ee
By (\ref{a9})--(\ref{a11}) we have 
\begin{eqnarray*}
&&\|[\phi]\|^2_t= ([\phi], [((t\s l)^2\D^2)^{-1}\phi_t])_{\D^2} \leq 
\|[\phi]\|_{\D^2}\|[((t\s l)^2\D^2)^{-1}\phi_t]\|_{\D^2} \\
&&=\|[\phi]\|_{\D^2}\sqrt{([\phi], [((t\s l)^2\D^2)^{-1}\phi_t])_t}\leq
\sqrt{c}\|[\phi]\|_{\D^2}\|[\phi]\|_t\, ,
\end{eqnarray*}
hence 
\be
\|[\phi]\|_t\leq \sqrt{c}\|[\phi]\|_{\D^2}\, ,
\label{a12}
\ee
and
\be
\left\|\lb((t\s l)^2\D^2)^{-1}\phi_t\rb\right\|_{\D^2}\leq 
c\|[\phi_t]\|_{\D^2}\, ,
\label{bd}
\ee
that is, the operator (\ref{ras}) is bounded with respect to this
norm as well. Furthermore, for any sequence $[\phi_n]\in[\Ci]$ there
is  
\begin{eqnarray*}
&&\|[\phi_n]\|^2_{\D^2}\leq \|[\phi_n-\phi_m]\|^2_{\D^2} +
2([\phi_n], [\phi_m])_{\D^2}\\ 
&&\leq\|[\phi_n-\phi_m]\|^2_{\D^2} +2\|[\phi_n]\|_t
\|[(t\s l)^2\D^2\phi_m]\|_t\, .
\end{eqnarray*}
Suppose $\dsp\lim_{m,n\to\infty}\|[\phi_m-\phi_n]\|_{\D^2}=0$ and 
$\dsp\lim_{n\to\infty}\|[\phi_n]\|_t=0$. For $\e>0$ let 
$\|[\phi_m-\phi_n]\|<\e$ for all $m,n\geq N$. Put $m=N$ and let
$N'\geq N$ be such that for all $n\geq N'$ there is 
$\dsp 2\|[\phi_n]\|_t\|[(t\s l)^2\D^2\phi_N]\|<\e^2$. Then for all
$n\geq N'$ there is $\|[\phi_n]\|^2_{\D^2}<2\e^2$, hence 
$\dsp\lim_{n\to\infty}\|[\phi_n]\|_{\D^2}=0$. Summing up, the norm
$\|.\|_{\D^2}$ is stronger than $\|.\|_t$ and these norms are
compatible \c{gelsh}. This implies that the Hilbert space $\H_{\D^2}$
obtained from $[\Ci]$ by completion with respect to $\|.\|_{\D^2}$ is
(may be canonically identified with) a subspace of the completion of
$[\Ci]$ with respect to $\|.\|_t$. The latter space is the Hilbert
space $L^2_t$ of equivalence classes of homogeneous of degree $0$, 
measurable functions modulo constant for which $\|[\phi]\|_t<\infty$.
This space does not depend on $t$ when considered as a vector space.
All elements of $\H_{\D^2}$ are therefore equivalence classes
$[\phi]$ of such functions. 

The operator (\ref{ras}) extends to a bounded, self-adjoint, positive
operator on $\H_{\D^2}$. In view of the comment following
(\ref{a7}) this is a compact operator with eigenvalues
$(j(j+1))^{-1}$ of multiplicity $2j+1$, where $j=1,2,\ldots$, and
eigenspaces $\H_j$ contained in $[\Ci]$. 

With the use of the Hilbert space $\H_{\D^2}$ the operation of
projecting $\H_0$ to $\HI$ needed in Sec.\ref{auw} may be
described more explicitly. Let, first, $f_a(l)$ be a smooth
function, homogeneous of degree $-1$ and orthogonal to $l^a$,
which means that $[f_a]$ is a smooth element of $\H_0$. For every
such function there are smooth, homogeneous of degree $0$ functions
$\phi(l)$ and $\p(l)$, unique up to additive constants, such that 
\be
l_af_b(l)-l_bf_a(l)=L_{ab}\phi(l)-{}^*L_{ab}\p(l)\, .
\label{a13}
\ee
This is shown most easily with the use of the spinor formalism, which
we do not intend to discuss here and refer the reader to \c{pr}, and
to \c{her95} for application to related problems. Within this
formalism it is a simple result that $\dsp o^{A'}f_{A'A}(l)=
\frac{\D}{\D o^A}\chi(l)$, where $o^A$ is the spinor of the null
vector $l^a$ and $\chi(l)$ is a smooth complex function, homogeneous
of degree $0$, determined by this equation up to an additive
constant. The equivalent form of this equation in the tensor language
is  ${}^-(l_af_b-l_bf_a)={}^-L_{ab}\chi$, where ${}^-G_{ab}$ for an 
antisymmetric tensor $G_{ab}$ denotes its antiselfdual part 
${}^-G_{ab}=\frac{1}{2}(G_{ab}+i{}^*G_{ab})$. Solving this equation
for $l_af_b-l_bf_a$ one obtains (\ref{a13}), with
$\phi=\mbox{Re}\chi$ and $\p=\mbox{Im}\chi$. Now, let $f_1$ and $f_2$
be two such smooth functions represented as in (\ref{a13}). Then by
(\ref{a13}) and the first equality in (\ref{a3}) 
\begin{eqnarray*}
&&f_1\s f_2(l)=(t\s l)^{-2}t^at^c(l_af_{1b}-l_bf_{1a}) 
(l_cf_2^b-l^bf_{2c})\\
&&=(t\s l)^{-2}\lp {}^*L_{0b}\phi_1{}^*L_0{}^b\phi_2 + 
{}^*L_{0b}\p_1{}^*L_0{}^b\p_2 - {}^*L_{0b}\p_1 L_0{}^b\phi_2 -
L_{0b}\phi_1{}^*L_0{}^b\p_2\rp\, .
\end{eqnarray*}
Now integrate this identity with respect to $\dl$, integrate one
${}^*L$ in each term by parts and use (\ref{a1}) and (\ref{a2}) to
obtain 
\be
([f_1], [f_2])_0=([\phi_1], [\phi_2])_{\D^2} +
([\p_1], [\p_2])_{\D^2} \, .
\label{a14}
\ee
Therefore, the map $(\phi, \p)\rightarrow f$ given by (\ref{a13}) for
smooth elements extends to a unitary map $\H_{\D^2}\dis \H_{\D^2}
\rightarrow \H_0$. The space $\HI$ is the image of the subspace 
$\H_{\D^2}\dis 0$ in this map.

\end{sloppypar}
\end{document}